\documentclass[10pt,twocolumn,superscriptaddress,english,10pt,prl,showpacs,floatfix,aps]{revtex4-1}
\usepackage[T1]{fontenc}           
\usepackage[utf8]{inputenc}
\usepackage[english,]{babel}
\usepackage{amsthm}
\usepackage{amsmath}
\usepackage{amssymb} 
\usepackage{esint}
\usepackage{graphicx}
\usepackage{upgreek}
\usepackage{xspace} 
\usepackage{nicefrac}
\usepackage{xcolor}
\usepackage{booktabs}
\usepackage{dcolumn}
\usepackage{notes2bib}

\makeatletter
\@ifundefined{textcolor}{}
{%
 \definecolor{BLACK}{gray}{0}
 \definecolor{WHITE}{gray}{1}
 \definecolor{RED}{rgb}{1,0,0}
 \definecolor{GREEN}{rgb}{0,1,0}
 \definecolor{BLUE}{rgb}{0,0,1}
 \definecolor{CYAN}{cmyk}{1,0,0,0}
 \definecolor{MAGENTA}{cmyk}{0,1,0,0}
 \definecolor{YELLOW}{cmyk}{0,0,1,0}
 \definecolor{ORANGE}{rgb}{1.0, 0.49, 0.0}
}

\usepackage{soul}
\usepackage{braket}
\usepackage[backref=false, 
bookmarksnumbered=true,
bookmarks=true,
bookmarksopen=true,
colorlinks=true,
citecolor=blue,
linkcolor=blue,
anchorcolor=green,
urlcolor=blue,unicode=false]{hyperref}
\usepackage{siunitx}

\renewcommand{\v}[1]{\ensuremath{\mathbf{#1}}} 
 
\let\baraccent=\= 
\renewcommand{\=}[1]{\stackrel{#1}{=}} 

\newcommand{\didv}{d$I/$d$V$\xspace}

\newcommand{\Fig}[1]{Fig.~\ref{#1}}
\newcommand{\Figure}[1]{Figure~\ref{fig:#1}}

\renewcommand{\Re}{\operatorname{Re}}

\def\iu{\ensuremath{\mathrm{i}}}

\makeatother

\usepackage{babel}

\begin{document}

\title{Photon-assisted tunneling at the atomic scale: Probing resonant Andreev reflections from Yu-Shiba-Rusinov states}

\author{Olof Peters}
\affiliation{\mbox{Fachbereich Physik, Freie Universit\"at Berlin, 14195 Berlin, Germany}}
\author{Nils Bogdanoff}
\affiliation{\mbox{Fachbereich Physik, Freie Universit\"at Berlin, 14195 Berlin, Germany}}

\author{Sergio Acero Gonzalez}
\affiliation{\mbox{Dahlem Center for Complex Quantum Systems and Fachbereich Physik, Freie Universit\"at Berlin, 14195 Berlin, Germany}}
\author{Larissa Melischek}
\affiliation{\mbox{Dahlem Center for Complex Quantum Systems and Fachbereich Physik, Freie Universit\"at Berlin, 14195 Berlin, Germany}}

\author{J. Rika Simon}
\affiliation{\mbox{Fachbereich Physik, Freie Universit\"at Berlin, 14195 Berlin, Germany}}
\author{Ga\"el Reecht}
\affiliation{\mbox{Fachbereich Physik, Freie Universit\"at Berlin, 14195 Berlin, Germany}}

\author{Clemens B. Winkelmann}
\affiliation{\mbox{Université Grenoble Alpes, CNRS, Institut Neél, 25 Avenue des Martyrs, 38042 Grenoble, France}}

\author{Felix von Oppen}
\affiliation{\mbox{Dahlem Center for Complex Quantum Systems and Fachbereich Physik, Freie Universit\"at Berlin, 14195 Berlin, Germany}}

\author{Katharina J. Franke}
\affiliation{\mbox{Fachbereich Physik, Freie Universit\"at Berlin, 14195 Berlin, Germany}}

\date{\today}

\begin{abstract}
Tunneling across superconducting junctions proceeds by a rich variety of processes, which transfer single electrons, Cooper pairs, or even larger numbers of electrons by multiple Andreev reflections. Photon-assisted tunneling combined with the venerable Tien-Gordon model has long been a powerful tool to identify tunneling processes between superconductors. Here, we probe superconducting tunnel junctions including an impurity-induced Yu-Shiba-Rusinov (YSR) state by exposing a scanning tunneling microscope with a superconducting tip to microwave radiation. We find that a simple Tien-Gordon description describes tunneling of single electrons and Cooper pairs into the bare substrate, but breaks down for tunneling via YSR states by resonant Andreev reflections. We develop an improved theoretical description which is in excellent agreement with the data. Our results establish photon-assisted tunneling as a powerful tool to analyze tunneling processes at the atomic scale which should be particularly informative for unconventional and topological superconductors.

\end{abstract}

\maketitle 
Subgap states in superconductors \cite{Balatsky2006,Heinrich2018,Sauls2018,Lutchyn2018} provide a local probe of interactions competing with superconductivity \cite{Deacon2010, Franke2011,Lee2014,Jellingard2016,Lee2017,Schneider2019,Liebhaber2020} and govern the Josephson coupling in superconducting junctions \cite{Bretheau2013}. They also provide promising arenas for quantum computing applications, owing to protection by the quasiparticle gap \cite{Zazunov2003,Nazarov2003, Janvier2015} or by topology \cite{Kitaev2001}. A wealth of information can thus be obtained from tunneling experiments on subgap states. The differential conductance reveals the energy of the bound states \cite{Yazdani1997, Ji2008}, spatially resolves electron and hole components of bound state wavefunctions in scanning tunneling microscopy \cite{Menard2015, Choi2017}, and provides access to inelastic relaxation rates underlying quasiparticle poisoning \cite{Ruby2015b,Albrecht2017}. In recent years, this has been instrumental in attempts to distinguish Majorana from Andreev and Yu-Shiba-Rusinov (YSR) states \cite{Lutchyn2018}.

Reliably extracting this information requires a microscopic understanding of the tunneling mechanism. Indeed, tunneling across superconducting junctions proceeds by a remarkably rich variety of processes. As single-electron tunneling requires sufficient energy to excite quasiparticles in the superconducting source and drain, transport at low energies is dominated by multi-electron processes including Cooper pair tunneling near zero bias \cite{Josephson1962, Namann2001} and multi-electron tunneling via (multiple) Andreev reflections at higher subgap biases \cite{Schrieffer1963, Taylor1963, Andreev1964, Ternes2006}. Unlike standard differential conductance measurements, photon-assisted tunneling directly reveals the effective charge of the transferred carriers through the sideband spacings in the bias voltage as well as in the amplitude of the high-frequency (HF) radiation \cite{Tien1963, Falci1991, Roychowdhury2015}. 

Here, we provide direct evidence for various tunneling mechanisms at the atomic scale by exposing a scanning tunneling microscope (STM) to high-frequency radiation. We first use photon-assisted tunneling from a superconducting tip into a pristine superconducting substrate to identify single-electron tunneling both above and within the superconducting gap, as well as Cooper-pair tunneling and Andreev reflections at subgap voltages.
We then exploit the resolution and the tunability of the junction conductance afforded by an STM to unravel charge transfer processes in the presence of atomic-scale structures \cite{Ruby2015b, Randeria2016, Farinacci2018, Brand2018}. We demonstrate this capability by investigating resonant tunneling via YSR states of individual magnetic adatoms. Remarkably, we observe that in the presence of subgap states, photon-assisted Andreev reflections exhibit distinct deviations from standard Tien-Gordon theory. Our more elaborate theory explains these deviations and shows how they encode properties of the subgap states as well as the underlying tunneling processes, including the transferred charge.

\begin{figure*}[tb]
	\includegraphics[]{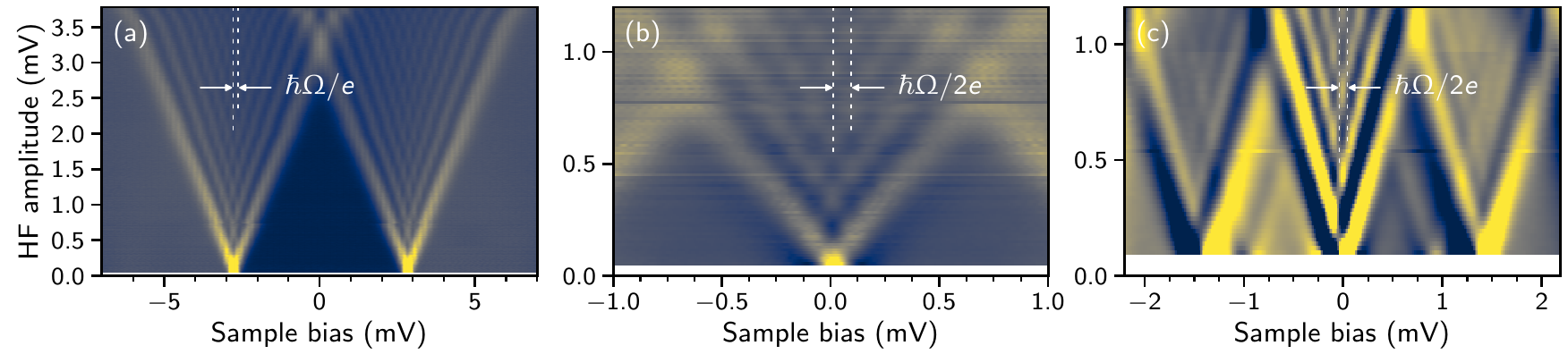} 
		 \caption{\label{Fig1} 
Color maps of \didv spectra of superconducting Pb-Pb junctions under HF radiation of $f=\SI{40}{GHz}$. 
(a) The BCS coherence peaks of a Pb--Pb(111) junction, framing the superconducting energy gap of $2\Delta$, split into V-shaped patterns under HF irradiation. The spacing of the sidebands corresponds to $\hbar \Omega /e$, reflecting single-electron tunneling (for a zoom on this splitting, see SI). 
(b) The Josephson peak at zero bias (recorded on a Pb(110) surface) splits into a V-shaped pattern with sideband spacing of $\hbar \Omega /2e$, reflecting Cooper-pair tunneling. 
(c) d$^2I$/d$V^2$ spectra around the first Andreev reflection at $eV=\pm\Delta$ split as $\hbar \Omega /2e$, reflecting two-electron transfer. The second derivative is taken for enhanced contrast. For parameters see SI.}
\end{figure*}

For our experiments, we implemented a high-frequency (HF) circuit into our STM setup at 1.3\,K with the final cable terminating close to the STM junction and acting as an efficient HF antenna [see Supplementary Information (SI)]. We use a superconducting Pb-coated tungsten tip facing a Pb(111) or Pb(110) surface. At large junction resistance, tunneling occurs beyond a threshold voltage of $\pm 2\Delta/e$, where single-electron tunneling becomes possible and excites one quasiparticle each in tip and substrate. In the presence of 40-GHz radiation, the resulting Bardeen-Cooper-Schrieffer (BCS) coherence peaks in the differential conductance \didv split into symmetric sidebands with distinct maxima spaced by $\SI{167(10)}{\mu V}$ (\Fig{Fig1}a). Electrons can emit or absorb $n$ photons while tunneling, so that the threshold voltages follow from $e|V| + n\hbar\Omega=2\Delta$. The photon sidebands are observable within a V-shaped region as a function of bias voltage and HF amplitude $V_\mathrm{HF}$ (\Fig{Fig1}a), as the number of photons $n<\Lambda =  eV_\mathrm{HF}/\hbar\Omega$ is limited by the maximal energy $eV_\mathrm{HF}$ which the tunneling electron can exchange with the radiation field. The separation $\hbar\Omega/e$ of sidebands is direct evidence for single-electron tunneling. 

A quantitative understanding is provided by the Tien-Gordon model which includes the HF radiation as an additional $ac$ voltage applied to the junction \cite{Tien1963}. Including the resulting modulation of the energy levels in the electrodes predicts differential conductance sidebands
\begin{equation}
\label{conductance}
G (V) = \sum_n J_n^2\left(\frac{keV_\mathrm{HF}}{\hbar\Omega}\right) G^{(0)} (V+n\hbar\Omega/k e).
\end{equation}
The strength of the $n$th sideband involves the Bessel function $J_n$ and $G^{(0)}$ denotes \didv in the absence of HF radiation. Here, we have already generalized the Tien-Gordon formula to the case that $k$ electrons are transferred in an elementary tunneling process \cite{Falci1991}. Using Eq.\ (\ref{conductance}) for single-electron tunneling ($k=1$) with the data in the absence of the HF radiation as input, we find excellent agreement with our experimental results (see SI). 

Unlike single-electron tunneling, Cooper-pair tunneling does not excite tip or substrate and can thus be observed as a zero-bias Josephson peak once the junction conductance is sufficiently large. The radiation field also splits the Josephson peak into a distinct V-shaped structure of resonant sidebands (\Fig{Fig1}b, see also \cite{Roychowdhury2015}). The sideband spacings in bias voltage and in HF amplitude are now $\SI{85(5)}{\mu V}$, corresponding to half the value observed for the BCS coherence peaks. This is consistent with $\hbar\Omega/2e$ as expected for electron pairs and thus provides direct evidence for Cooper-pair tunneling. Indeed, simulations based on Eq.~(\ref{conductance}) with $k=2$ are in excellent agreement with the experimental maps (see SI). 

\begin{figure}[tb]
	\includegraphics[]{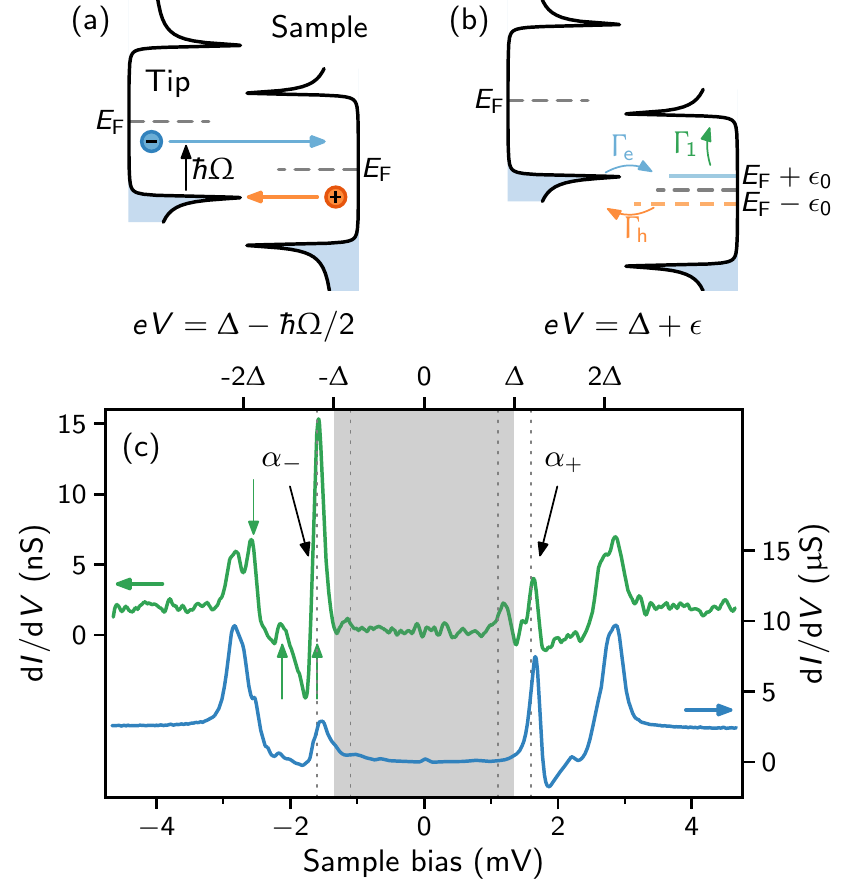} 
		\caption{\label{Fig2} (a) Photon-assisted tunneling process leading to the first sideband of the first Andreev reflection at bias voltage $eV=\Delta-\hbar \Omega/2e$. 
		 (b) Sketch of tunneling processes with YSR states (without HF radiation): $\Gamma_e$ transfers an electron from tip to YSR state, which relaxes by $\Gamma_1$ into the continuum (single-electron tunneling) or is reflected as a hole with rate $\Gamma_h$ (resonant Andreev reflection).  (c) \didv spectra recorded with a superconducting Pb tip on a Mn atom adsorbed on a Pb(111) surface: green spectrum in single-electron tunneling regime; blue spectrum in resonant Andreev regime, parameters in SI.
		Green arrows mark the YSR resonances at negative biases. The lowest-lying YSR state at $e{V}=\pm(\Delta+\epsilon_0)$ (indicated by dashed lines) is marked by $\alpha_\pm$. The shaded area indicates the superconducting gap of the tip.
}
		
\end{figure}

To further substantiate our ability to determine the elementary charge carriers in superconducting STM junctions, we investigate Andreev reflections in the HF field \cite{Chauvin2006}. The first Andreev reflection can be resolved at $eV=\pm\Delta$ for sufficiently high junction conductance ($G_\mathrm{N}=\SI{40}{\mu S}$  in \Fig{Fig1}c). This \didv peak emerges from the transfer of a Cooper pair into the substrate,  leaving two quasiparticles behind in the tip, and a corresponding process generating quasiparticles in the substrate. The radiation field splits the Andreev reflection peaks into V-shaped patterns similar to those of the Josephson peak. (Note that we show $\mathrm{d}^2I/\mathrm{d}V^2$ for better resolution.) We find a sideband spacing of $\SI{91(9)}{\mu V}$, equal to $\hbar\Omega/2e$, revealing the underlying charge transfer of $2e$ (see \Fig{Fig2}a for a sketch of the process leading to the first sideband). 

The Tien-Gordon approach as embodied by Eq.~(\ref{conductance}) is thus highly successful at describing photon-assisted tunneling between tip and bulk superconducting substrate, directly revealing the number of electrons involved in the elementary tunneling process. We now extend the technique to tunneling into YSR states induced by magnetic adatoms. Remarkably, we find that in general, Eq.~(\ref{conductance}) breaks down for tunneling into subgap states and has to be replaced by a more elaborate description. 

We induce YSR states by depositing Mn atoms on a Pb(111) surface. Mn atoms have been shown to remain stable when approached by the tip, with the junction conductance changing over several orders of magnitude \cite{Ruby2015b}. Differential conductance spectra recorded on the Mn atoms reveal several YSR states (\Fig{Fig2}c, blue) at bias voltages $eV=\pm(\Delta_\mathrm{tip}+\epsilon)$, with YSR-state energies $\epsilon =$ \SIlist{0.25;0.77;1.2}{meV}. These can be understood as a multiplet of $d$ orbitals
with its degeneracy partially lifted by the local crystal field \cite{Ruby2016}. Here, we focus on the most intense YSR resonance at $\epsilon_0 =$ \SI{0.25}{meV}. While the position of the resonance, labeled as $\alpha$, remains unchanged when approaching the adatom with the STM tip and changing the junction conductance by three orders of magnitude, the relative magnitude of the resonances at positive and negative bias changes and eventually inverts (see \Fig{Fig2}c and extracted intensities in SI). The inversion reflects a transition from dominant single-electron tunneling at large tip distances (\Fig{Fig2}c, green) to dominant Andreev processes at small distances (\Fig{Fig2}c, blue) and manifests the asymmetry between the electron wave function $u$ and hole wavefunction $v$ of the YSR state, $|u|^2\ll|v|^2$ \cite{Ruby2015b}. 

\begin{figure*}[tb]
	\includegraphics[]{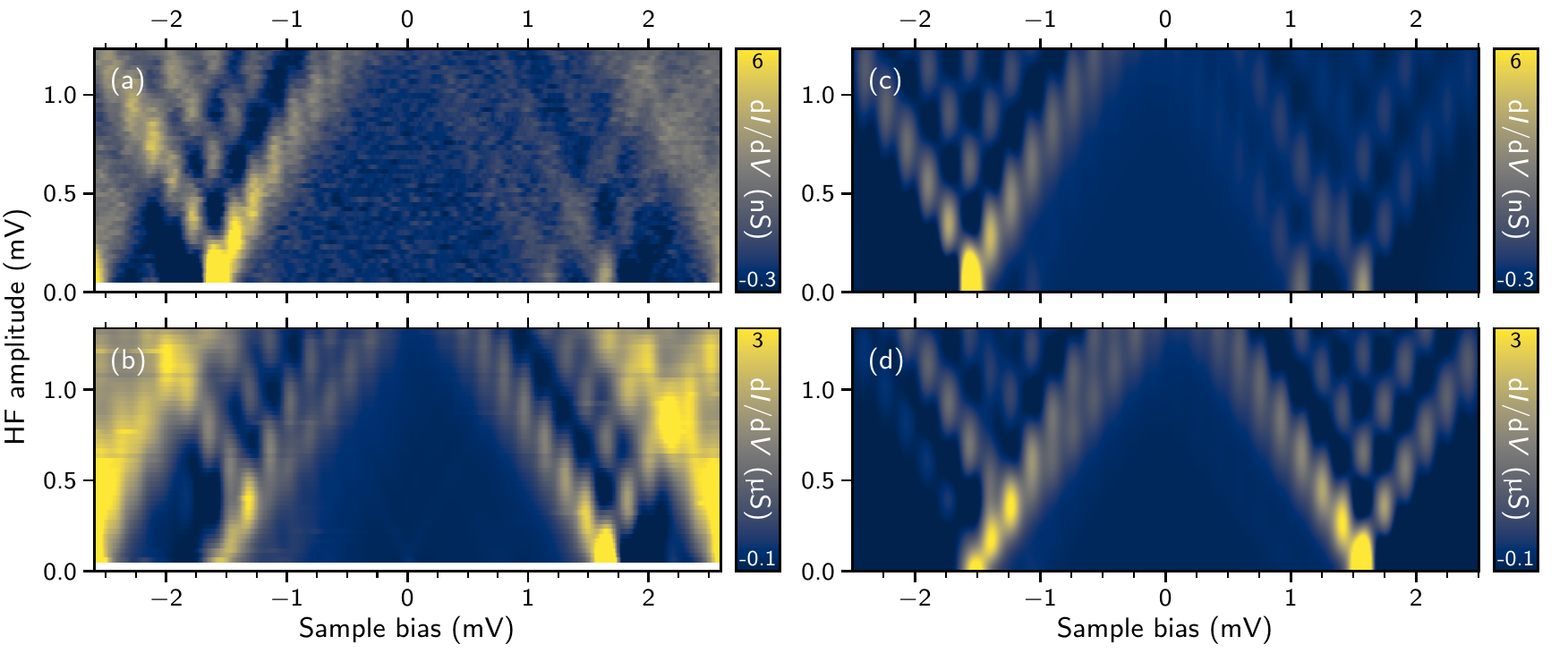} 
		\caption{\label{Fig3} Photon-assisted tunneling into YSR state under HF irradiation: (a) V-shaped splittings of the YSR peaks for a low (normal-state) junction conductance ($G_\mathrm{N}^\text{low}=\SI{2.6e-5}{}G_0$ with ${G_0}=2e^2/h$). The faint structure emerging from $eV=\Delta-\epsilon_0 = \SI{1.1}{mV} $ is due to thermally excited quasiparticles.
(b) At high conductance ($G_\mathrm{N}^\text{high}=\SI{5.2e-2}{G_0}$), we observe a V-shaped pattern at positive bias and a distinct Y-shaped pattern with a double-peak structure on the low-bias branch at negative bias.
(c, d) Simulations corresponding to the junction conductances in (a, b) are in excellent agreement with experiment (BCS peaks omitted for clarity). For parameters, see SI.}
\end{figure*}

Measured \didv maps of photon-assisted tunneling through Pb--Mn--Pb junctions are shown in \Fig{Fig3}a and b and exhibit 
clear deviations from the Tien-Gordon expression in Eq.~(\ref{conductance}).
At high junction conductances (\Fig{Fig3}b), the maps still resemble a V-shaped pattern for positive bias voltages, but are distinctly different for negative biases. Most strikingly, the V-shaped pattern gives way to a Y-shaped form, which resembles none of the maps for the unstructured surface. Even for positive bias voltages with its V-shaped pattern, the data are surprising. While tunneling at these junction conductances is dominated by Andreev reflections and thus transfers electron pairs \cite{Ruby2015b}, the periods of the intensity modulations seemingly indicate single-electron tunneling. 

These unexpected results contrast with \didv maps at lower junction conductance (\Fig{Fig3}a), which show V-shaped patterns centered at $e|V|=\Delta+\epsilon_0$ for both bias polarities. Moreover, the observed period of the intensity modulations is consistent with the expectation \cite{Ruby2015b} that tunneling is dominated by individual electrons which leave behind a quasiparticle in the tip and occupy the YSR state at energy $\epsilon_0$ [see \Fig{Fig2}b, with a similar hole process for $eV=-(\Delta+\epsilon_0$)]. With the corresponding thresholds $eV +n\hbar\Omega = \Delta + \epsilon_0$ for photon-assisted tunneling, we apply the Tien-Gordon model to the individual tunneling rates and obtain
\begin{align}
	\Gamma_{e,\text{eff}}(E)= \sum_{n} J_n^2\left(\frac{e V_\text{HF}}{\hbar \Omega}\right) \Gamma_{e}(E+n \hbar \Omega)
\label{gammae}
\end{align}
for electrons of energy $E$ in the substrate. Here, $\Gamma_{e}(E)$ is the tunneling rate in the absence of radiation (see SI for details). At these low junction conductances, inelastic processes in the substrate (with rate $\Gamma_1$, see \Fig{Fig2}b) empty the YSR state long before the next tunneling event, so that tunneling is the rate-limiting step. The current can then be computed by integrating the tunneling rate (\ref{gammae}) and a corresponding hole rate $\Gamma_{h,\text{eff}}(E)$ with appropriate thermal occupation factors. (For the full expressions, see SI.) The result in \Fig{Fig3}c exhibits excellent agreement with our data in \Fig{Fig3}a. In particular, the calculation also reproduces a second V-shaped structure centered at $eV = \Delta - \epsilon_0$ which originates from thermally activated quasiparticles  \cite{Ruby2015b} (see also SI). Unlike our simulations for unstructured substrates, these simulations use only the YSR asymmetry and relaxation rates as derived from experiment as input, providing strong confirmation for the interpretation of the underlying processes. 

\begin{figure}[tb]
	\includegraphics[]{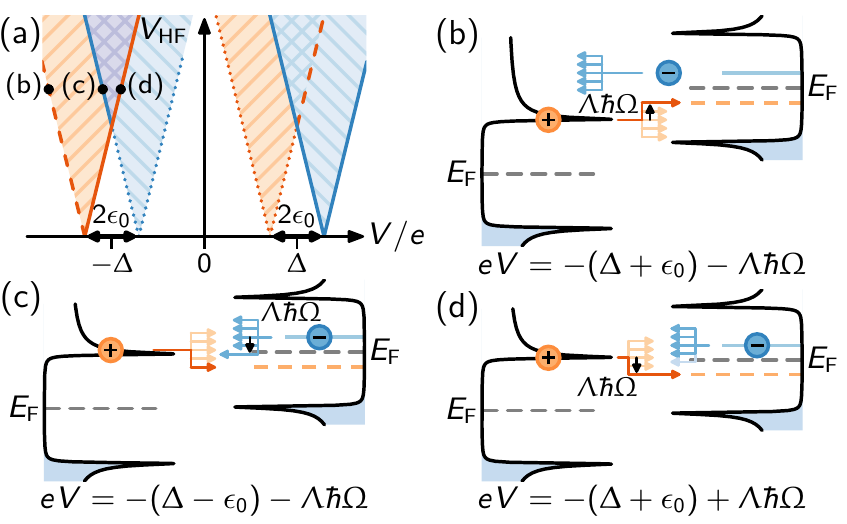} 
		\caption{\label{Fig4} Photon-assisted resonant Andreev reflections with YSR state: (a) V-shaped regions with separate photon-assisted tunneling thresholds for electrons (blue; emerging from $V=\pm\Delta+\epsilon_0$) and holes (red; $V=\pm\Delta-\epsilon_0$). Both electron and hole tunneling must be above threshold, so that structure is only seen in outer V shapes. A large asymmetry between electron and hole YSR wavefunctions makes photon sidebands most visible in the (blue) electron regions. At negative bias, this highlights the purple region, resulting in a pronounced Y shape, as indicated by full lines. At positive bias, the outer V (blue) is due to electron tunneling and sidebands are visible throughout. Dots marked (b)-(d) indicate bias voltages and HF amplitude for which tunneling processes are sketched in corresponding panels. These show resonant Andreev reflections with: (b) On-threshold (i.e., from just outside the BCS gap) hole tunneling emitting the maximal number $\Lambda=eV_{\mathrm{HF}}/\hbar\Omega$ of photons with above-threshold electron tunneling. (c) On-threshold electron tunneling emitting the maximal number $\Lambda$ of photons, with hole processes partially above threshold. (d) On-threshold hole tunneling absorbing the maximal number $\Lambda$ of photons, electron processes partially above threshold. On-threshold processes are indicated by long arrows, processes above (below) threshold by short full (faint) arrows.
}
\end{figure}

At larger junction conductances, inelastic relaxation out of the YSR state becomes slower than electron and hole tunneling across the junction. Tunneling is then dominated by resonant Andreev reflections. Focusing first on negative bias voltages near $eV=-(\Delta+\epsilon_0)$, this process involves hole tunneling into the negative-energy YSR state, electron tunneling out of the positive energy YSR state, and breaking of a Cooper pair in the substrate (\Fig{Fig2}b). The resonant nature of this process has two important implications. First, Andreev reflection from YSR states is much stronger than from unstructured substrates. Second, there are {\em separate} thresholds for electron and hole tunneling. The condition $eV +m\hbar\Omega = -(\Delta + \epsilon_0)$ for holes is complemented by $eV +n\hbar\Omega = -(\Delta - \epsilon_0)$ for electrons (see Fig.\ \ref{Fig4} for illustration and representative processes). Without HF radiation, the electron process is automatically above threshold when the hole process is. But with HF radiation, this is in general no longer the case. 

As a result of the maximal number of absorbed or emitted photons, the separate thresholds define two distinct V-shaped regions at negative biases, one for holes, $|eV+(\Delta + \epsilon_0)|<eV_{\mathrm{HF}}$ (left red V in \Fig{Fig4}a), and another one for electrons, $|eV+(\Delta - \epsilon_0)|<eV_{\mathrm{HF}}$ (left blue V in \Fig{Fig4}a). Resonant Andreev reflection takes place when electron and hole processes are both above threshold, so that only the outer (red) V-shaped region appears in experiment. The Y shape results from the strongly asymmetric electron and hole amplitudes of the YSR state, $|u|^2\ll |v|^2$. Due to this asymmetry, the hole tunneling rate $\Gamma_{h,\text{eff}}$ is generically significantly larger than the electron rate $\Gamma_{e,\text{eff}}$. Thus, electron tunneling is, roughly speaking, the rate-limiting step governing the tunneling current, and features due to $\Gamma_{e,\text{eff}}$ are much more pronounced. This explains why sidebands appear with high intensity in regions where the two V shapes overlap and why the observed maxima remain spaced by $\hbar\Omega/e$ despite the underlying two-electron tunneling. The thresholds for the hole processes still contribute to the low-voltage arm of the Y (full red line in \Fig{Fig4}a). Here, hole tunneling just sets in and is comparable in rate to electron tunneling. Thus, in this region, hole thresholds are visible in addition to electron thresholds, resulting in the prominent double structure of the low-voltage arm of the Y shape observed in \Fig{Fig3}b (see SI for more details). The results of a full theory of resonant Andreev reflections, which properly treats it as a single coherent process involving both electron and hole tunneling, are shown in \Fig{Fig3}d. Using only few experimental parameters as input, we find excellent agreement with our data, reproducing the Y shape including the double structure as well as the sideband spacing $\hbar\Omega/e$ (see SI for details). 

At positive bias, the outer V shape, centered at $eV=\Delta+\epsilon_0$ (blue in \Fig{Fig4}a), emerges from the electron condition, and the inner one around $eV=\Delta-\epsilon_0$ (red) from the hole condition. Again, structure is only visible within the outer (blue) V shape where both electron and hole processes are above threshold. But the `rate-limiting' electron tunneling is now associated with the outer (blue) V shape. Consequently, pronounced sidebands appear within the entire outer V shape, again spaced by $\hbar\Omega/e$. These observations are also nicely reproduced by our theoretical calculations of the resonant Andreev reflections in \Fig{Fig3}d. We finally emphasize that the inner V shapes at large junction conductance should not be confused with the thermal processes at low junction conductance. Thermal processes require inelastic relaxation and are therefore much weaker than resonant Andreev reflections at large junction conductance. 

In conclusion, exposing a scanning tunneling microscope with a superconducting tip to high-frequency radiation constitutes a powerful tool to investigate the subgap structure of superconducting substrates with atomic resolution. Here, we have established this ability for YSR states and showed that the data provide a comprehensive picture of the tunneling process. Exploiting the tunability of the junction conductance over several orders of magnitude, we have specifically obtained exquisite insight into the contributions of Andreev processes to subgap tunneling. 

Such Andreev processes should provide a particularly revealing window into the physics of unconventional and topological superconductors, making atomic-resolution photon-assisted tunneling highly promising for such systems. For instance, strongly disordered superconducting films are believed to exhibit a transition from a pseudogap regime with preformed Cooper pairs into a superconducting phase with long-range phase coherence \cite{Dubouchet2019}. Spatially resolved photon-assisted tunneling promises detailed information on the nature of subgap tunneling processes, their spatial variation, and hence the nature of pairing correlations. As a second example, resonant Andreev reflections underlie tunneling experiments probing putative Majorana modes \cite{Law2009,Flensberg2010}, for instance in chains of magnetic adatoms on $s$-wave superconductors \cite{NadjPerge2014,Ruby2015, Pawlak2016, Feldman2017, Jeon2017,Kim2019}. Photon-assisted tunneling can not only ascertain the nature of the tunneling processes, but also serve to distinguish true zero-energy states from states with a small, but finite energy $\epsilon_0$. In fact, finite-energy states with electron and hole wave functions of the same order exhibit splittings due to the separate electron and hole thresholds which should be detectable with high resolution due to their large multiplicity throughout the $V$-shaped region.

\begin{acknowledgments}
We thank K. Flensberg for fruitful discussions and C. Lotze for technical support. We gratefully acknowledge funding by the European Research Council under the Consolidator Grant "NanoSpin", by Deutsche Forschungsgemeinschaft and Agence National de Recherche under grant "JOSPEC", and by CRCs 183 and 910 of Deutsche Forschungsgemeinschaft.  
\end{acknowledgments}

\bibliographystyle{apsrev4-1}

\clearpage

\setcounter{figure}{0}
\setcounter{section}{0}
\setcounter{equation}{0}
\setcounter{table}{0}
\renewcommand{\theequation}{S\arabic{equation}}
\renewcommand{\thefigure}{S\arabic{figure}}
	\renewcommand{\thetable}{S\arabic{table}}%
	\setcounter{section}{0}
	\renewcommand{\thesection}{S\arabic{section}}%

\onecolumngrid


\renewcommand{\Fig}[1]{Fig.~\ref{fig:#1}}
\renewcommand{\Figure}[1]{Figure~\ref{fig:#1}}

\newcommand{\vsigma}{\mbox{\boldmath $\sigma$}}

\section*{\Large{Supplementary Information}}

\section{Experimental Details}
\subsection{Implementation of high-frequency circuit}
Exposing a scanning tunneling microscope (STM) to high-frequency (HF)  radiation in the GHz range requires a dedicated circuit \cite{SBaumann2015, SNatterer2019,SFriedlein2019,SSeifert2020}. Typical cables in low-temperature STMs are designed for efficient filtering of environmental radiation, which would otherwise induce noise in the tunneling junction. Thus, high-frequency signals require separate cables, which transmit the signal without compromising other stability and resolution criteria of the STM.

\begin{figure}[b]
	\includegraphics[]{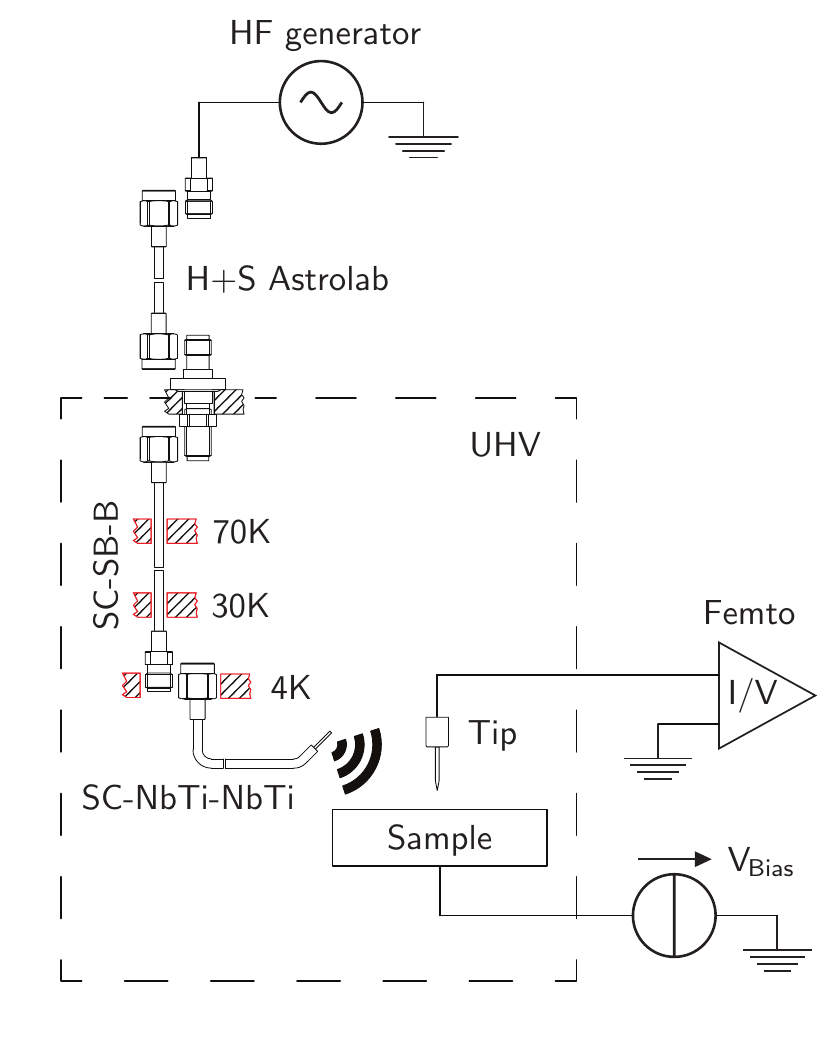} 
	\caption{\label{fig:RF-Cabeling} Sketch of the HF circuit used in the experiments. The HF signal is fed into the UHV chamber via HF cables, anchored at various stages for thermalization. The antenna close to the STM junction is formed by the open end of the coaxial cable. For details see text.
	} 
\end{figure}

We have designed a circuit in parallel to the STM cables. A sketch of the setup is shown in \Fig{RF-Cabeling}. The HF signal from a signal generator (R\&S\textsuperscript{\textregistered} SMB 100A) is transmitted to the chamber via a semi-flexible cable (H+S Astrolab KK-SF240-2X11SK, $l=\SI{0.5}{\metre}$). All connectors outside and inside the chamber as well as the feedthrough (Allectra 242-SMAD40G-C16) are designed as SMK \SI{2.92}{\milli\metre}, suited for frequencies up to \SI{40}{\giga\hertz}. Inside the UHV chamber, a silver-plated beryllium-copper coaxial cable (SC-219/50-SB-B, $l=\SI{1.25}{\metre}$) is passed through the cryostat and attached to the cryostat's radiation shields. At the bottom of the \SI{4}{\kelvin}-plate of the croystat it is connected (via SMK connectors) to a superconducting niobium titanate (NbTi) coaxial cable (SC-086/50-NbTi-NbTi, $l=\SI{0.4}{\metre}$), which has negligible attenuation losses at low temperature. This cable terminates close to the STM junction with the outer conductor being removed over a length of $\SI{5}{\milli\metre}\approx\lambda_\text{\SI{30}{\giga\hertz}}/2$ and bent up \SI{\sim 45}{\degree}. 

\subsection{Characterization of HF circuit}

The HF signal applied to the antenna effectively modulates the bias voltage across the STM junction. When we applied the minimum output signal, i.e., \SI{-20}{dBm} of our signal generator (R\&S\textsuperscript{\textregistered} SMB 100A), the superconducting energy gap completely vanished, reflecting an effective bias voltage modulation of several mV. For lower bias-voltage modulations, we inserted a \SI{40}{\deci \bel} attenuator at the output of the signal generator. To characterize the HF radiation arriving at the STM junction, we make use of the highly nonlinear current-voltage characteristics of a superconductor-superconductor tunneling junction. A similar scheme can be used at other nonlinearities such as inelastic spin excitations \cite{SBaumann2015}.

\begin{figure}[tb]
	\includegraphics[]{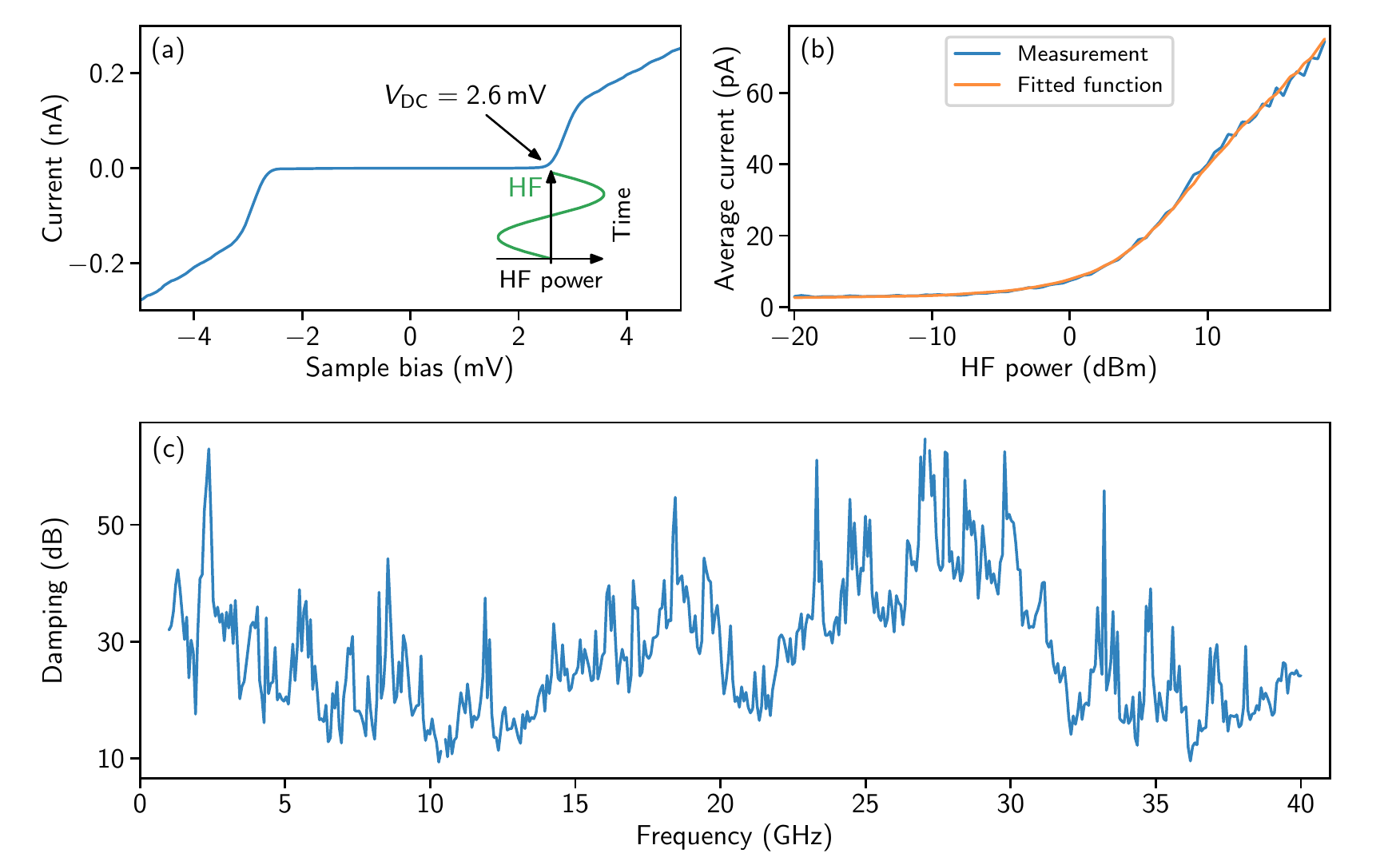} 
	\caption{\label{fig:HF_Characterization} Transmission characteristics of HF circuit. (a) $I(V)$ characteristic of a Pb--Pb tunnel junction at $G_\text{N}=\SI{3.7e-1}{G_0}$. 
	For measuring the transmission characteristics, the $dc$ bias voltage is set just below the superconducting energy gap while the HF signal is applied to the antenna. The HF modulation of the sample bias voltage leads to a time-averaged current, which depends on the effective $ac$ amplitude at the junction. (b) The dependence of the time-averaged current on applied HF power is fitted by Eq.\ (\ref{RF}) to determine the effective power arriving at the STM junction. 
(c) The frequency-dependent damping is extracted from fits as shown in (b).}
\end{figure}

We set the $dc$ bias voltage $V_\mathrm{dc}$ close to the nonlinearity in the current-voltage characteristic $I(V)$. In the case of a Pb--Pb junction, we applied $V=\SI{2.6}{\milli \volt}$, which is just below the onset of quasiparticle tunneling across the superconductor--superconductor junction at $eV=2\Delta$. We then switch on the HF radiation (see sketch in \Fig{HF_Characterization}a) and measure the time-averaged current $\left\langle I \right\rangle$ as a function of applied HF power (\Fig{HF_Characterization}b). We determine the damping by fitting the experimental time-averaged current to
\begin{equation}
\label{RF}
\left\langle I(V_\text{dc}) \right\rangle = \frac{1}{2\pi} \int_0^{2\pi} I \left[\,V_\mathrm{dc}+\sin(\Phi) \sqrt{2 R \cdot 10^{\frac{L(\text{dBm})-g(\text{dB})}{10}}\SI{1}{\milli \watt}}\, \right]d\Phi
\end{equation}
where $g(\text{dB})$ is the damping, $L(\text{dBm})$ is the power of the source in dBm and $R=\SI{50}{\ohm}$ is its impedance. Here, we use that the power level $L$ in dBm can be converted to conventional units by $P$(mW)= $10^{\frac{L}{10}}$\,mW. The extracted damping is shown in \Fig{HF_Characterization}c.

Alternatively, the effective radiation power at the STM junction can be determined by comparison with Tien-Gordon theory. The $V$-shaped splitting of, e.g., the BCS coherence peaks in the power-dependent \didv maps provides an accurate measure of the radiation power that leads to photon-assisted tunneling (see main text). This method is very efficient for specific frequencies, but would be rather time consuming for a full frequency-dependent characterization of the HF~circuit. We used the method to monitor variations of the damping properties over the timescale of the measurements for the presented experiments at a constant frequency of \SI{40}{\giga \hertz}. 

\subsection{Experimental parameters}

Table \ref{tab:SpecParamters} collects the experimental parameters at which the measurements presented in the figures, both in the main text and the supplementary, were taken. 

\setlength{\tabcolsep}{0.1em}
\begin{table}[h]
	\caption{Compilation of experimental parameters used for recording the \didv spectra in main text and supplement} 
	\label{tab:SpecParamters}
	\small 
	\centering 
	\begin{tabular}{lddddr} 
	\toprule\toprule
	 & V_\text{Bias}\,\text{[mV]} & I\,\text{[nA]} & \Delta z\,\text{[pm]} & V_\text{LockIn}\,[\upmu \text{V}_\text{RMS}] & $f_\text{LockIn}\,\text{[Hz]}$ \\ 
	\midrule
	Fig. 1a; \Fig{ExperimentVsSimulation}a; \Fig{ExperimentVsSimulation}c 	& 10 	& 0.5 	& 0 	& 40 	& 873 	\\
	Fig. 1b; \Fig{ExperimentVsSimulation}e 									& 10 	& 50 	& 0 	& 40 	& 873 	\\
	Fig. 1c; \Fig{ExperimentVsSimulation}g 									& 5 	& 200 	& 0 	& 20 	& 929 	\\
	Fig. 2c (green); Fig. 3a; \Fig{PeakForPeak}a  							& 5 	& 0.1 	& -100 	& 40 	& 929 	\\
	Fig. 2c (blue); Fig. 3b; \Fig{PeakForPeak}b  							& 5 	& 20 	& 0 	& 20 	& 929 	\\
	\Fig{HF_Characterization}a; \Fig{HF_Characterization}b 					& 10 	& 0.287	& 0 	& - 	& $-$ 	\\
	\bottomrule
	\end{tabular}
 \end{table}

\section{Theoretical considerations}

Here, we present theoretical considerations on photon-assisted tunneling in superconductor-superconductor junctions. We begin with a brief summary of photon-assisted tunneling from a superconducting tip into a pristine superconducting substrate, including the three processes for which we show experimental data in the main text: (i) single-electron tunneling at $e|V|\simeq 2\Delta$ (coherence peaks), (ii) Andreev reflections at $e|V|\simeq\Delta$, and (iii) the Josephson peak at $e|V|\simeq 0$. We will use a Fermi-golden-rule approach for all three of these processes. This then provides the background for a discussion of photon-assisted resonant Andreev reflections through Yu-Shiba-Rusinov (YSR) states. 

\subsection{General formulation}

Tunneling from a superconducting tip with Hamiltonian $\tilde H_{\rm tip}$ into a superconducting substrate with Hamiltonian $\tilde H_{\rm sub}$ is described by the tunneling Hamiltonian
\begin{equation}
\tilde H_T=\sum_\sigma\left[t{c}^\dagger_{{\rm tip},\sigma}({\bf R}) {c}_{{\rm sub},\sigma}({\bf R})+{\rm h.c.}\right],
\end{equation}
where ${c}^\dagger_{\alpha,\sigma}({\bf r})$ denotes the electron creation operators at position ${\bf r}$ and with spin $\sigma$ in the tip ($\alpha={\rm tip}$) or the substrate ($\alpha={\rm sub}$). The position of the tip is denoted by ${\bf R}$. The overall Hamiltonian 
\begin{equation}
   \tilde H = \tilde H_{\rm tip} + \tilde H_{\rm sub} + \tilde H_T
\label{hambef}
\end{equation}
conserves the total number of particles $N=N_{\rm tip}+N_{\rm sub}$. The bias voltage $V(\tau)$ between tip and substrate is included through different chemical potentials $\mu_{\rm tip}$ and $\mu_{\rm sub}$ for tip and substrate, 
\begin{equation}
   eV(\tau) = \mu_{\rm tip}-\mu_{\rm sub},
\end{equation} 
and consists of an applied $dc$ voltage $V$ as well as an $ac$ voltage, 
\begin{equation}
      V(\tau) =  V + V_{\rm HF}\cos(\Omega\tau).
\end{equation}
The $ac$ voltage is generated by the radiation field of frequency $\Omega$ \cite{STien1963}. 

The BCS descriptions of the superconducting tip and substrate start from the grand-canonical Hamiltonians $H_{\rm tip} = \tilde H_{\rm tip} - \mu_{\rm tip} N_{\rm tip}$ and $H_{\rm sub} = \tilde H_{\rm sub} - \mu_{\rm sub} N_{\rm sub}$, which measure the single-particle energies in tip and substrate from the respective chemical potentials $\mu_{\rm tip}$ and $\mu_{\rm sub}$. This makes it useful to perform the time-dependent canonical transformation 
\begin{equation}
   U(\tau) = \exp\left\{ \frac{i}{\hbar} \int_0^{\tau} {\mathrm d}\tau^\prime \left[\mu_{\rm tip}\left(\tau^\prime\right) N_{\rm tip} + \mu_{\rm sub}\left(\tau^\prime\right) N_{\rm sub}\right]  \right\},
\end{equation}
which transforms the Hamiltonian (\ref{hambef}) into $H=U\tilde H U^\dagger-i\hbar U\partial_\tau U^\dagger$, i.e., 
\begin{equation}
   H = \left(\tilde H_{\rm tip} -\mu_{\rm tip} N_{\rm tip}\right) + \left(\tilde H_{\rm sub} - \mu_{\rm sub} N_{\rm sub}\right) + U\tilde H_T U^\dagger.   
\end{equation}
Here we used that $\tilde H_{\rm tip}$ and $\tilde H_{\rm sub}$ conserve $N_{\rm tip}$ and $N_{\rm sub}$, so that $U \tilde H_\alpha U^\dagger = \tilde H_\alpha$. In contrast, the tunneling Hamiltonian only conserves the total number of electrons $N$, but not $N_{\rm tip}$ and $N_{\rm sub}$ separately. This makes the transformed tunneling Hamiltonian $H_T = U\tilde H_T U^\dagger$ time dependent,
\begin{equation}
H_T=\sum_\sigma\left[te^{-i\phi(\tau)}{c}^\dagger_{{\rm sub},\sigma}({\bf R}){c}_{{\rm tip},\sigma}({\bf R})+{\rm h.c.}\right],
\label{tdeptun}
\end{equation}
where the time-dependent phase
\begin{equation}
   {\phi(\tau)}= \frac{1}{\hbar}\int_0^\tau {\mathrm d}\tau' eV(\tau') =   \frac{eV}{\hbar}\tau+\frac{eV_{\rm HF}}{\hbar\Omega}\sin(\Omega\tau)
\end{equation} 
includes the effects of the applied bias voltage. 

The transformed Hamiltonian can now be treated within BCS mean field approximation, so that the unperturbed Hamiltonian becomes
\begin{eqnarray}
H_0=H_{\rm tip} + H_{\rm sub}=\sum_{{\bf k},\alpha} \sum_{\sigma} \left[\xi_{{\bf k},\alpha} {c}^\dagger_{\alpha,\mathbf{k}\sigma}{c}_{\alpha,\mathbf{k}\sigma}+\left(\Delta{c}^\dagger_{\alpha,\mathbf{k}\uparrow}{c}^\dagger_{\alpha,-\mathbf{k}\downarrow}+{\rm h.c.}\right) \right]. 
\end{eqnarray}
Here, $\xi_{{\bf k},\alpha}=\epsilon_{\bf k}-\mu_\alpha$ denotes the normal-state dispersion, the superconducting gap $\Delta$ is taken to be identical for tip and substrate, and we temporarily assume a pristine substrate without magnetic adatom. 

\subsection{Tunneling into pristine superconductors}

\subsubsection{Coherence peaks}

Single-electron tunneling across a superconducting junction creates a quasiparticle each in source and drain electrode in the final state and consequently requires a threshold voltage of $2\Delta/e$. The resulting coherence peaks are split into sidebands by the radiation field. To obtain these sidebands, we expand the time-dependent phase factor appearing in the tunneling Hamiltonian into a Fourier series,
\begin{equation}
   e^{-i\phi(\tau)} = \sum_n J_n\!\left( \frac{eV_\mathrm{HF}}{\hbar\Omega} \right) e^{-ieV\tau/\hbar -in\Omega\tau}  
\end{equation}
with the coefficients involving the Bessel functions $J_n(x)$. Assuming that temperature is small compared to the superconducting gap, tunneling only occurs from source to drain, and Fermi's golden rule gives 
\begin{equation}
    I = 2 e \sum_n J^2_n\!\left( \frac{eV_\mathrm{HF}}{\hbar\Omega} \right)  \sum_{\bf k} \sum_{\bf k'} \frac{2\pi}{\hbar} |t|^2 u_{\bf k}^2 v_{\bf k'}^2 \delta(E_{\bf k} + E_{\bf k'} - eV -n\hbar\Omega) 
\label{sbset}
\end{equation}
for the tunneling current. Here, we have expressed the electron operators ${c}^\dagger_{\alpha,\sigma}$ in terms of Bogoliubov quasiparticle operators in the usual way, $u_{\bf k}$ and $v_{\bf k'}$ denote the familiar electron and hole quasiparticle wavefunctions in BCS theory, and $E_{\bf k} =  [\xi_{\bf k}^2 + \Delta^2]^{1/2}$ is the quasiparticle energy. (We assume identical superconductors for tip and substrate, as is the case in the experiment.) The prefactor of two accounts for spin. Thus, we obtain \cite{STien1963}
\begin{equation}
   I(V) = \sum_n J^2_n\!\left( \frac{eV_\mathrm{HF}}{\hbar\Omega} \right)  I_0(V +n\hbar\Omega/e)
\label{TGcoherence}
\end{equation}
in terms of the current-voltage characteristic $I_0(V)$ in the absence of the radiation field. A standard calculation reduces $I_0(V)$ to the familiar result \cite{STinkham2004}
\begin{equation}
 I_0(V) = \frac{4\pi e |t|^2}{\hbar} \int \mathrm{d}E \, \nu_{\mathrm {tip}}(E-eV)\nu_{\mathrm {sub}}(E) [n_\text{F}(E-eV)- n_\text{F}(E)] 
\end{equation}
of the semiconductor model. Here, $n_\text{F}(E)$ denotes the Fermi function and 
\begin{equation}
	\label{eq:SC_DOS_Theory}
   \nu(E)=\nu_{\mathrm {tip}}(E) = \nu_{\mathrm {sub}}(E) = \nu_0 \frac{|E| \, \theta\left(|E|-\Delta\right)}{\sqrt{E^2 -\Delta^2}} 
\end{equation}
the (identical) superconducting densities of states of tip and substrate. ($\nu_0$ is the normal state density of states.)

Equation (\ref{sbset}) shows that the sidebands of the coherence peaks are spaced by $\hbar\Omega/e$ in $dc$ bias voltage $V$. The oscillations of the Bessel functions $J_n(x)$ for $x>n$ imply modulations of the sideband strength with a period of $\hbar\Omega/e$ as a function of the amplitude $V_{\mathrm {HF}}$ of the radiation field. Finally, the Bessel functions are strongly suppressed for $x<n$, limiting the sidebands to a $V$-shaped region $V_{\mathrm{HF}} > |V|-2\Delta/e$ in color plots as a function of $V$ and $V_{\mathrm{HF}}$.

\subsubsection{Andreev reflections}

At subgap voltages and sufficiently low temperatures, the elementary tunneling processes involve the transfer of multiple electrons. Here, we focus on Andreev processes transfering two electrons across the junction which are still clearly resolved in our experiment. This process creates a Cooper pair in the drain electrode, leaving behind two quasiparticles in the source, or annihilates a Cooper pair in the source, creating two quasiparticles in the drain. Each of the two tunneling electrons gains an energy of $eV$ due to the applied $dc$ bias \footnote{{One can equivalently consider the tunneling process as energy conserving (in the absence of the radiation field) with different chemical potentials of source and drain, or the chemical potentials as identical, but the tunneling process as changing the electron energy by $\pm eV$. We base pictorial representations of the tunneling processes on the first possibility. The second possibility is adapted to the time-dependent tunneling Hamiltonian (\ref{tdeptun}), and we use corresponding language here when discussing the theoretical calculations.}} so that the process requires a threshold voltage $\Delta/e$. The corresponding amplitude can be obtained by expanding the $T$-matrix to quadratic order in the tunneling Hamiltonian. The intermediate state has two quasiparticles, one each in the source and drain electrodes, and the tunneling electron gains an energy of order $\Delta$ due to the applied bias voltage. Thus, the intermediate state has an energy denominator of order $\Delta$ in the vicinity of the threshold voltage and in the absence of the HF radiation. The $ac$ bias changes the energy denominator by at most $eV_{\rm {HF}}$ due to photon emission and absorption. As long as $eV_{\rm {HF}}$ is small compared to $\Delta$, we can approximate the second-order contribution $H_T G_0 H_T$ to the $T$ matrix as
\begin{equation}
   T \simeq  -H_T\frac{1}{\Delta} H_T.
\end{equation}
Then, the time-dependent phase factors $e^{-i\phi(\tau)}$ from both tunneling Hamiltonians combine into 
\begin{equation}
   e^{-2i\phi(\tau)} = \sum_n J_n\!\left( \frac{2eV_\mathrm{HF}}{\hbar\Omega} \right) e^{-i2eV\tau/\hbar -in\Omega\tau}.
\end{equation}
A Fermi golden rule calculation of the contribution of Andreev reflection to the current will thus give sidebands (see \cite{SZimmermann1996} for an alternative Blonder-Tinkham-Klapwijk approach), 
\begin{equation}
   I(V) = \sum_n J^2_n\!\left( \frac{2eV_\mathrm{HF}}{\hbar\Omega} \right)  I_0(V +n\hbar\Omega/2e),
\label{TGAndreev}
\end{equation}
where $I_0(V)$ is the current-voltage characteristic in the absence of the radiation field. This Tien-Gordon-like formula directly encodes the two-electron nature of the tunneling process through the sideband spacing $\hbar\Omega/2e$ in bias voltage and the oscillation period $\hbar\Omega/2e$ as a function of $V_{\mathrm{HF}}$. This derivation also implies that deviations from the Tien-Gordon-like formula can appear for larger HF amplitudes or larger HF frequencies. For the parameters of our experiment, these deviations appear weak.

\subsubsection{Josephson peak}

In our experiment, the tunneling current near zero bias is due to incoherent transfer of Cooper pairs between source and drain electrodes. In the absence of the radiation field, the Cooper pairs gain or lose an energy of $2eV$ in the tunneling process. Cooper pair tunneling from the source to the drain electrode must therefore be associated with dissipating excess energy to the electromagnetic environment of the junction (see, e.g., \cite{SIngold1992} for a review). (A minimal model of the environment is an Ohmic resistor $R$ in series with the junction \cite{SIngold1992}.) Conversely, tunneling of Cooper pairs from drain to source electrode can occur when absorbing an energy of $2eV$ from the environment. Within a strictly classical description of the environment, absorption and (stimulated) emission are equally likely, resulting in a vanishing net current across the junction. A nonzero net current  appears in a quantum description of the environment due to spontaneous emission processes \footnote{The Josephson peak can also be described within the RCSJ model, expressing the Josephson current in terms of the phase difference across the junction and computing the (Langevin) dynamics of the phase to linear order in the Josephson energy. While one obtains the same result, this calculation requires one to include only classical voltage fluctuations.}.

Cooper-pair transfer proceeds via an intermediate state with a quasiparticle each in source and drain electrode. Since the Josephson peak occurs for $eV\ll\Delta$, the corresponding energy denominator can be approximated by $2\Delta$. Again, as long as $eV_{\rm {HF}}\ll 2\Delta$, the energy denominator remains unaffected by photon absorption or emission. Then, as for Andreev reflections, the time-dependent phase factors combine and the amplitude for Cooper-pair transfer from tip to substrate becomes 
\begin{equation}
 \frac{E_\text{J}}{2} e^{- i2\phi(\tau)} = \frac{E_\text{J}}{2} e^{-i2\tilde\phi}\sum_n J_n\!\left( \frac{2eV_\mathrm{HF}}{\hbar\Omega} \right) e^{- i2eV\tau/\hbar - in\Omega\tau}
\end{equation}
with the Josephson energy $E_\text{J}$ (obeying the usual Ambegaokar-Baratoff relation). To describe the coupling to the environment, the tunneling Hamiltonian needs to be complemented by the operator $e^{-i2\tilde\phi}$ describing the charge transfer of $2e$ associated with the tunneling Cooper pair. Fermi's golden rule then yields the Tien-Gordon-like expression for the current (see \cite{SFalci1991} for a functional integral approach)
\begin{equation}
    I(V) = \sum_n J^2_n\!\left( \frac{2eV_\mathrm{HF}}{\hbar\Omega}\right) I_0(V+n\hbar\Omega/2e),
   \label{TGJosephson}
\end{equation}
which directly encodes the transfer of electron pairs between tip and substrate. The current in the absence of the HF radiation takes the form \cite{SIngold1992}
\begin{equation}
    I_0(V) = 2e \left(\frac{E_\text{J}}{2\hbar}\right)^2 2\pi\hbar \left[P(2V) - P(-2V) \right]
\end{equation}
and 
\begin{equation}
   P(E)  = \frac{1}{2\pi\hbar} \int {\mathrm d}t e^{iE\tau/\hbar + J(\tau)}.
\end{equation}
Neglecting the junction capacitance for simplicity, the function $J(\tau)$ in the exponent takes the form \cite{SIngold1992}
\begin{equation}
   J(\tau) = 2\int\frac{\mathrm{d}\omega}{\omega} \frac{R}{R_Q} \left\{\coth \frac{\beta\hbar\omega}{2} [\cos\omega\tau -1] -
     i\sin\omega\tau\right\},
\end{equation}
where $R_Q = h/4e^2$. While the first term in the curly brackets is symmetric in $\tau$ and thus classical for $\hbar\omega\ll T$, the second term is odd and purely quantum. Expanding to linear order in this term and performing the integrals, one finds the result \cite{SIvanchenko1969,SGrabert1999,SNamann2001}
\begin{equation}
    I_0(V) = \frac{I_c^2R}{2}\frac{V}{V^2 + \left( \frac{2e}{\hbar}RT \right)^2}
\end{equation} 
for the Josephson peak in terms of the critical current $I_c = 2e E_\text{J}/\hbar$ of the junction. 

\subsection{Resonant Andreev reflections through YSR states}

Like Andreev reflections and Josephson tunneling into pristine superconductors, resonant Andreev reflections through YSR states also transfer electron pairs between substrate and tip. Nevertheless, HF radiation affects resonant Andreev reflections in qualitatively different ways compared to the tunneling processes into pristine superconductors discussed above. The simple Tien-Gordon-like expressions (\ref{TGAndreev}) and (\ref{TGJosephson}) for photon-assisted tunneling of electron pairs used in essential ways that the energy denominator of the intermediate state and hence the pair tunneling amplitude can be taken as independent of energy (and thus a $\delta$-function in time). It is a result of this fact that the time-dependent phase factors associated with the two tunneling electrons combine into a single factor which in effect halves the spacing between sidebands compared to single-electron tunneling. Clearly, this breaks down when considering resonant Andreev reflections through a bound state of energy $\epsilon_0$ for which the pair tunneling amplitude has a sharp resonance associated with the bound state. 

Moreover, tunneling into pristine superconductors could be described in low-order perturbation theory in the tunneling Hamiltonian. In general, this is no longer possible for resonant Andreev reflections. In fact, the rates for tunneling across the junction can also dominate the width of the bound-state resonance, thus necessitating a treatment to all orders in perturbation theory \cite{SRuby2015b}. In experiment, the width is dominated by inelastic relaxation processes for small tunneling amplitudes (large tip-substrate distances) and by tip-substrate tunneling for large tunneling amplitudes (small tip-substrate distances) \cite{SRuby2015b}. In the first case, the current is due to single-electron transfers. An electron tunnels across the junction into the YSR state and subsequently, the resulting quasiparticle is inelastically excited into the continuum. This process can be described in low-order perturbation theory in the tunneling Hamiltonian. In the second case, the transfer of the first electron is accompanied by the transfer of a second electron, with the two electrons combining into a Cooper pair in the substrate. Alternatively, this process can be described within the Andreev picture in which electron tunneling is reflected as a hole. The width of the bound state resonance is now dominated by the electron and hole tunneling rates, so that the description of this process must include all orders in perturbation theory. 

\subsubsection{Resonant Andreev reflections in the absence of a radiation field}

In general, the tunneling current via YSR states is the sum of single- and two-electron (resonant Andreev) processes, $I = I_s + I_a$. Reference \cite{SRuby2015b} derived the expressions
\begin{alignat}{2}
I_s(V) &= \frac{e}{h} &\int  \mathrm{d}\omega &\frac{\Gamma_1 \left[ \Gamma_e^{n_\text{F}}(\omega) - \Gamma_h^{n_\text{F}}(\omega)  \right] - \Gamma_2 \left[ \Gamma_e^{1-n_\text{F}}(\omega) - \Gamma_h^{1-n_\text{F}}(\omega)  \right]  }{(\omega - \epsilon_0)^2 + \Gamma(\omega)^2/4}, \label{eq:Is} \\
I_a(V) &= \frac{2e}{h} &\int  \mathrm{d}\omega &\frac{\Gamma_h(\omega)\Gamma_e^{n_\text{F}}(\omega) - \Gamma_e(\omega)\Gamma_h^{n_\text{F}}(\omega)  }{(\omega - \epsilon_0)^2 + \Gamma(\omega)^2/4}\label{eq:Ia}
\end{alignat}
for these currents in the absence of HF irradiation. These expressions clearly display the resonance structure of the tunneling amplitudes as a result of the YSR state of energy $\epsilon_0$. The width $\Gamma(\omega) = \Gamma_1 + \Gamma_2 + \Gamma_{e}(\omega)+ \Gamma_{h}(\omega)$ of the resonance includes the tunneling rates  
\begin{alignat}{1}
\Gamma_{e}(\omega) &= 2\pi t^2 |u|^2 \nu(\omega_-), \\
\Gamma_{h}(\omega) &= 2\pi t^2 |v|^2 \nu(\omega_+) 
\end{alignat}
for electrons and holes as well as the rates $\Gamma_1$ and $\Gamma_2$ for inelastic excitations. Here, $\Gamma_1$ denotes the rate for exciting a quasiparticle occupying the positive-energy YSR state into the quasiparticle continuum, and $\Gamma_2$ the rate for occupying the YSR state from the continuum. We take $\Gamma_1$ and $\Gamma_2$ as phenomenological parameters of the model, which can be extracted directly from experiment outside the regime of resonant Andreev reflections. In the expressions for the electron and hole tunneling rates, $u$ and $v$ are the electron and hole wave functions of the YSR state at the location of the tip and $\omega_\pm = \omega \pm eV$. We also introduced the notation
\begin{alignat}{1}
\Gamma_{e}^{n_\text{F}}(\omega) &= 2\pi t^2 |u|^2 \nu(\omega_-)n_\text{F}(\omega_-), \\
\Gamma_{e}^{1-n_\text{F}}(\omega) &= 2\pi t^2 |u|^2 \nu(\omega_-)[1-n_\text{F}(\omega_-)].
\end{alignat}
The corresponding definitions for $\Gamma_{h}^{n_\text{F}}$ and $\Gamma_{h}^{1-n_\text{F}}$ just differ by replacing $|u|^2 \rightarrow |v|^2$ and $\omega_- \rightarrow \omega_+$. 

Experimentally, the regimes of dominant single-electron and dominant Andreev current can be distinguished by the dependence of the current on the normal-state conductance of the tunnel junctions \cite{SRuby2015b}. While the single-electron current is proportional to the normal-state conductance, the Andreev current has a sublinear dependence as a result of the tunneling rates appearing in the denominator of the bound-state resonance. The two regimes are also characterized by different asymmetries between the peak heights of the positive- and negative-voltage resonant Andreev peaks at $eV = \pm (\Delta + \epsilon_0)$. Analyzing the peak conductances at positive and negative bias \cite{SRuby2015b}, one finds that in the regime of dominant single-electron tunneling, their ratio is equal to $|u/v|^2$. This uses the fact that the broadening $\Gamma(\omega)$ is dominated by the relaxation rates $\Gamma_1$ and $\Gamma_2$. In the regime of dominant resonant Andreev processes, the broadening is dominated by the tunneling rates $\Gamma_e(\omega) +\Gamma_h(\omega)$, leading to an inversion of the ratio of peak heights which is now equal to  $|v/u|^{10/3}$.

\subsubsection{Photon-assisted resonant Andreev reflections through YSR states}

It is natural to expect that HF radiation modifies the tunneling rates for electrons and holes due to photon emission and absorption processes. As the rates describe tunneling of single electrons and holes, we assume that the corresponding sideband spacing is equal to $\hbar\Omega$ and that the weights of the sidebands are again given in terms of Bessel functions. We thus make the replacements 
\begin{alignat}{1}
\Gamma_{e}(\omega)         &= 2\pi t^2 |u|^2 \sum_n J_n^2 \left( {\frac{eV_\mathrm{HF}}{\hbar\Omega}} \right) \nu (\omega_{-n}), \\
\Gamma_{e}^{n_\text{F}}(\omega)   &= 2\pi t^2 |u|^2 \sum_n J_n^2 \left( {\frac{eV_\mathrm{HF}}{\hbar\Omega}} \right) \nu (\omega_{-n})n_\text{F}(\omega_{-n}), \\
\Gamma_{e}^{1-n_\text{F}}(\omega) &= 2\pi t^2 |u|^2 \sum_n J_n^2 \left( {\frac{eV_\mathrm{HF}}{\hbar\Omega}} \right) \nu (\omega_{-n})[1-n_\text{F}(\omega_{-n}) ]
\end{alignat}
in Eqs.\ (\ref{eq:Is}) and (\ref{eq:Ia}), where we defined $\omega_{\pm n} = \omega \pm eV \pm n\Omega$. The corresponding definitions for $\Gamma_{h}$ replace $|u|^2 \rightarrow |v|^2$ and $\omega_{-n} \rightarrow \omega_{+n}$. In this form, the rates $\Gamma_{e,h}(\omega)$ account for the electron and hole tunneling rates and resulting broadening, including absorption and emission of any number of photons. Our theoretical simulations of photon-assisted tunneling into YSR states are based on these expressions combined with the current equations (\ref{eq:Is}) and (\ref{eq:Ia}). We note that it is possible to give a systematic derivation of these expressions for photon-assisted tunneling through YSR states in the framework of a Keldysh Green function approach. This is beyond the scope of the present paper and will be presented in a separate publication. 

\begin{figure}[tb]
	\includegraphics[]{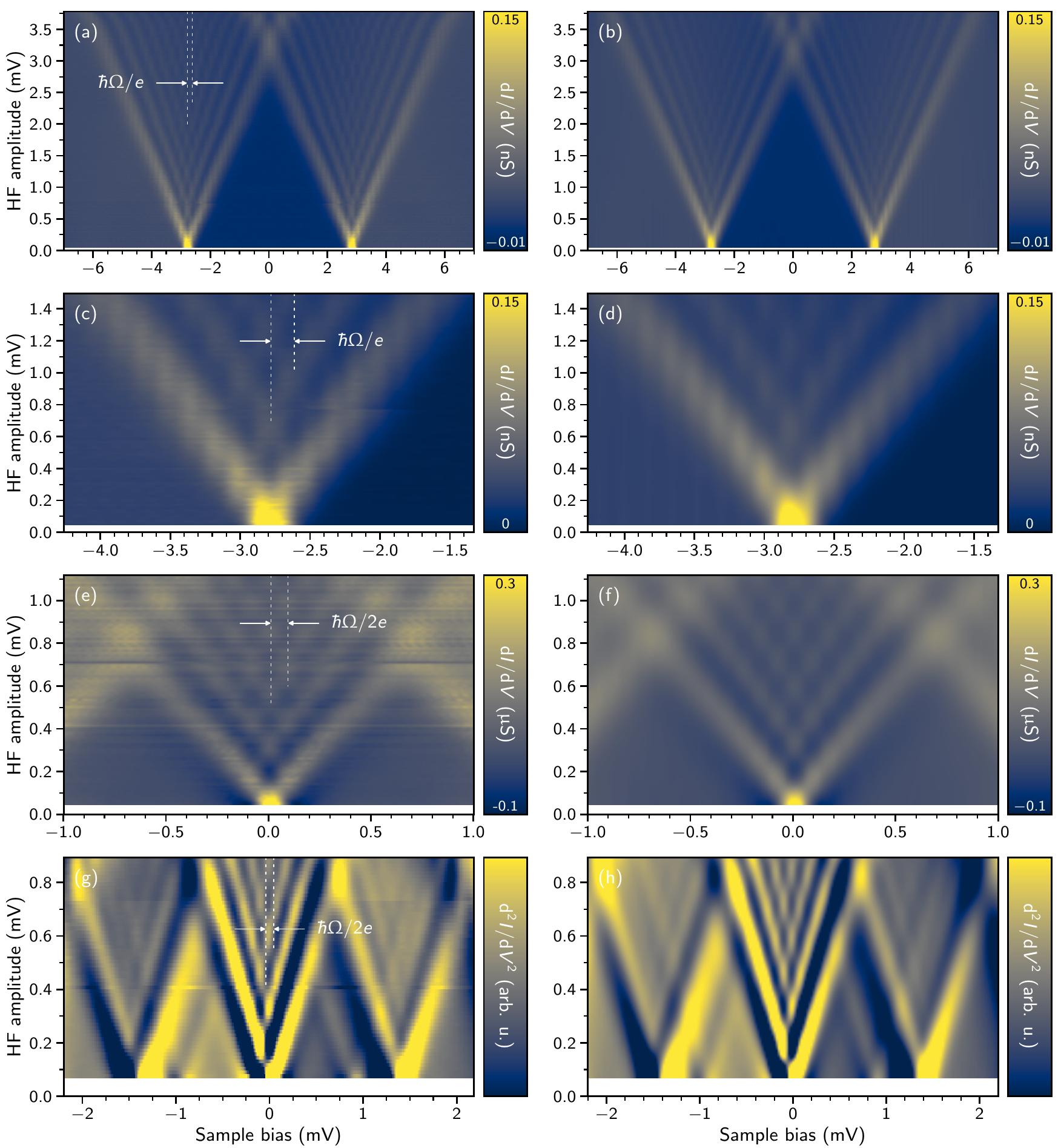} 
	\caption{\label{fig:ExperimentVsSimulation} Comparison of experimental (left) and simulated (right) \didv spectra under 40-GHz radiation. (a, b) V-shaped splitting of the BCS coherence peaks in a Pb--Pb junction. (c, d) Close-up views on the V shapes at low radiation power clearly reveal the energy splitting by $\hbar \Omega /e$. 
	(e, f) Splitting of the Josephson peak at zero bias (recorded on a Pb(110) surface)  with sideband spacing of $\hbar \Omega /2e$, reflecting Cooper-pair tunneling.
	(g, h) d$^2I$/d$V^2$ spectra around first-order multiple Andreev reflections at $eV=\pm\Delta$ split as $\hbar \Omega /2e$, reflecting two-electron transfer in a single tunneling event. The second derivative is taken for enhanced contrast. 
	The simulations are based on the Tien-Gordon-like expressions in Eqs.\ (\ref{TGcoherence}), (\ref{TGAndreev}), and (\ref{TGJosephson}) with an applied HF voltage of $V_\mathrm{HF}$, a frequency $\Omega= 2 \pi \SI{40}{GHz}$, and $G^{(0)}(V)$ as the measured spectrum without applied HF. For the one-electron processes in (a-d) $k=1$, and for the two-electron processes in (e)-(g) $k=2$. For experimental parameters, see table \ref{tab:SpecParamters}. }
\end{figure}

\section{Further comparison of experiment and theory}

\subsection{Simulations of photon-assisted tunneling into pristine substrates}

We first consider photon-assisted tunneling into a pristine substrate. Eqs.\ (\ref{TGcoherence}), (\ref{TGAndreev}), and (\ref{TGJosephson}) give Tien-Gordon-like expressions for the tunneling current, expressing the current in the presence of HF radiation in terms of the current without irradiation. Using these expressions, we can simulate the data in the presence of the HF irradiation based on our data without irradiation. The only free parameter is the damping determining the effective HF voltage $V_\mathrm{HF}$ at the junction.

\Figure{ExperimentVsSimulation} shows that corresponding simulations of photon-assisted tunneling are in excellent agreement with our data for all three relevant processes, namely the Josephson peak in \Fig{ExperimentVsSimulation}a,b [Eq.\ (\ref{TGJosephson})], Andreev reflection in \Fig{ExperimentVsSimulation}c,d [Eq.\ (\ref{TGAndreev})], and the coherence peaks in \Fig{ExperimentVsSimulation}e,h [Eq.\ (\ref{TGcoherence})]. In all cases, one correctly reproduces the $V$-like splitting of the peaks as well as the dependence of the sidebands on bias voltage and HF amplitude. 

We note that in the experiment, we artificially increased the damping of the transmission line with a \SI{\sim 40}{dB} attenuator. We observe that the total attenuation slowly fluctuates between \SIrange{53}{57}{dB}. We attribute these fluctuations to external influences such as the He level of the cryostat. For the duration of one measurement sweep, however, we find the attenuation to be constant to a good approximation. We determine the damping by fitting one of the spectra of each sweep using the appropriate Tien-Gordon-like expressions in Eqs.\ (\ref{TGcoherence}), (\ref{TGAndreev}), and (\ref{TGJosephson}).

\begin{figure}[tb]
	\includegraphics[]{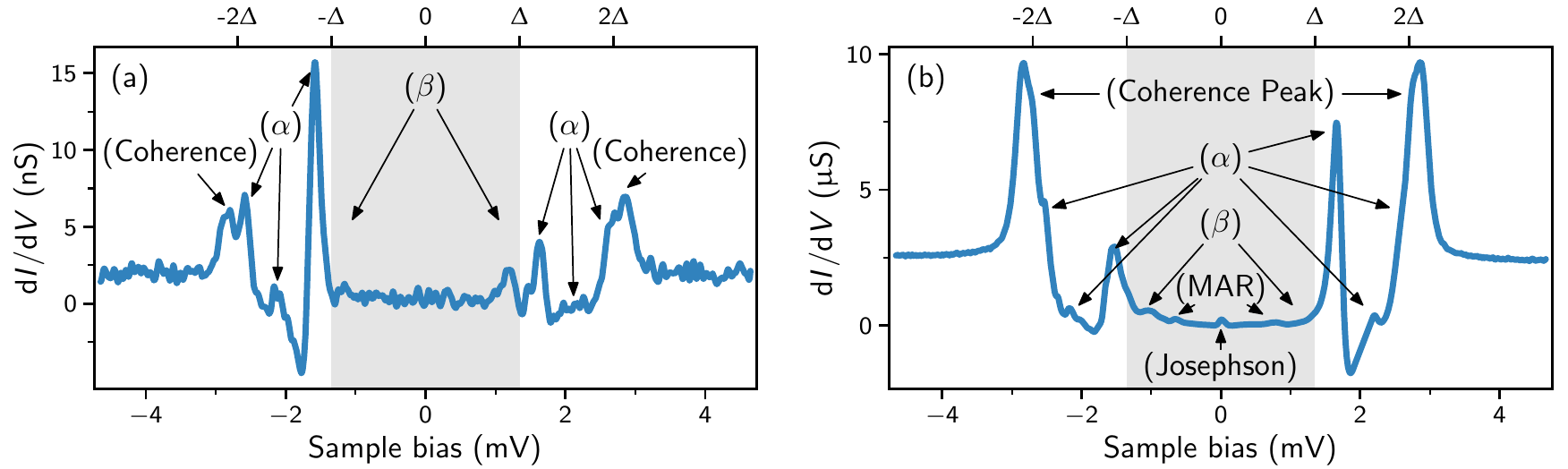} 
	\caption{\label{fig:PeakForPeak} \didv spectra recorded above the center of a single Mn adatom on Pb(111) at two tip-substrate distances: (a) Low conductance $G_\text{N}^\text{low}=\SI{2.6e-5}{G_0}$ 
	and (b) high conductance $G_\text{N}^\text{high}=\SI{5.2e-2}{G_0}$. 
The spectra exhibit the following distinctive features from high to low bias voltages: The coherence peaks are followed by three YSR resonances (marked by $\alpha$).  The tip gap is shown as gray shaded areas. Tunneling of thermally excited quasiparticles via YSR states is marked by $\beta$. At high junction conductance -- panel (b) -- we also observe multiple (resonant) Andreev reflections, marked by MAR, and Josephson tunneling.}
\end{figure} 

\newpage 
\subsection{Basic tunneling processes in the presence of YSR states}

\subsubsection{Tunneling processes without HF radiation}

\begin{figure}[tbh]
	\includegraphics[]{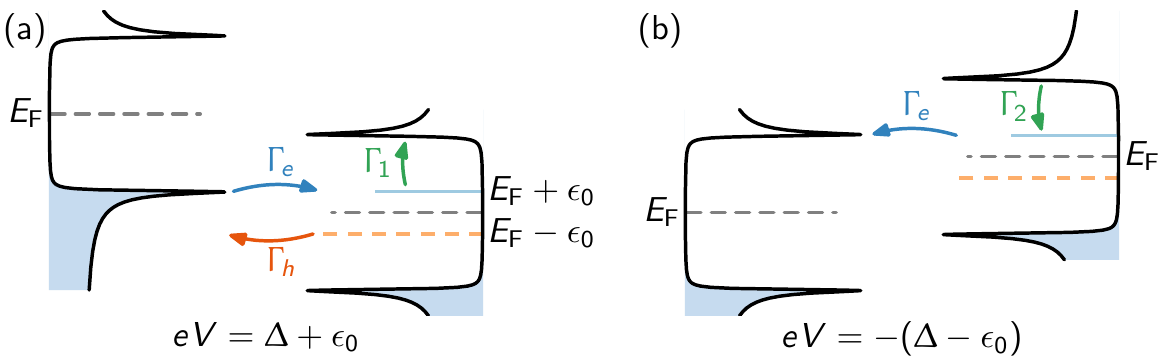} 
	\caption{\label{fig:Thermal_Gamma} (a) Sketch of tunneling processes via YSR state: At $eV= \Delta + \epsilon_0$, single-electron tunneling transfers an electron with rate $\Gamma_e$ into the positive-energy YSR state, which is then excited with rate $\Gamma_1$ into the positive-energy quasiparticle continuum of the substrate. Resonant Andreev reflections transfer a Cooper pair to the substrate via electron tunneling with rate $\Gamma_e$ and hole tunneling with rate $\Gamma_h$. (b) Thermal occupation of the YSR state with rate $\Gamma_2$ followed by single-electron tunneling with rate $\Gamma_e$ leads to a thermal replica of the experimental YSR resonance at $eV= -(\Delta - \epsilon_0)$. The analogous hole processes take place at $eV= -(\Delta + \epsilon_0)$ for panel (a), and $eV= +(\Delta - \epsilon_0)$ for panel (b). 
	}
\end{figure}

We now turn to our data on tunneling in the presence of YSR states. The \didv traces in the absence of HF radiation in \Figure{PeakForPeak} give an overview of the observed basic tunneling processes. Panel (a) corresponds to a low junction conductance (large tip-sample distance). We observe three peaks at $eV=\pm(\Delta+\epsilon)$ labeled by $\alpha$, which we associate with three YSR states at energies $\epsilon =$ \SIlist{0.25;0.77;1.2}{meV} induced by the Mn adatom. At small junction conductance, tunneling into the YSR states at $eV=\Delta+\epsilon$ is dominated by single-electron tunneling with rate $\Gamma_e$, followed by inelastic excitation with rate $\Gamma_1$ into the quasiparticle continuum of the substrate (see \Fig{Thermal_Gamma}a for a sketch). Similarly, at $eV=-(\Delta+\epsilon)$, tunneling proceeds by a corresponding hole process in which a hole tunnels from the tip into the negative-energy substrate YSR state, followed by a relaxation process with rate $\Gamma_2$ which refills the negative energy YSR state \cite{SRuby2015b}. 

For small junction conductances, there are additional resonances at $eV=\pm(\Delta-\epsilon_0)$ labeled by $\beta$ \cite{SRuby2015b}. These resonances are weaker compared to the resonances at $eV=\pm(\Delta+\epsilon_0)$ since they involve thermally excited quasiparticles. For electron tunneling, the process requires thermal occupation of the positive-energy YSR state with rate $\Gamma_2$, before tunneling into the tip with rate $\Gamma_e$ (see \Fig{Thermal_Gamma}b for a sketch). This leads to a thermal replica of the YSR state at $eV= -(\Delta - \epsilon_0)$. The corresponding hole process is found at $eV= (\Delta - \epsilon_0)$.

The single-electron tunneling processes are described theoretically by the expression given in Eq.\ (\ref{eq:Is}). 

\begin{figure}[tb]
	\includegraphics[]{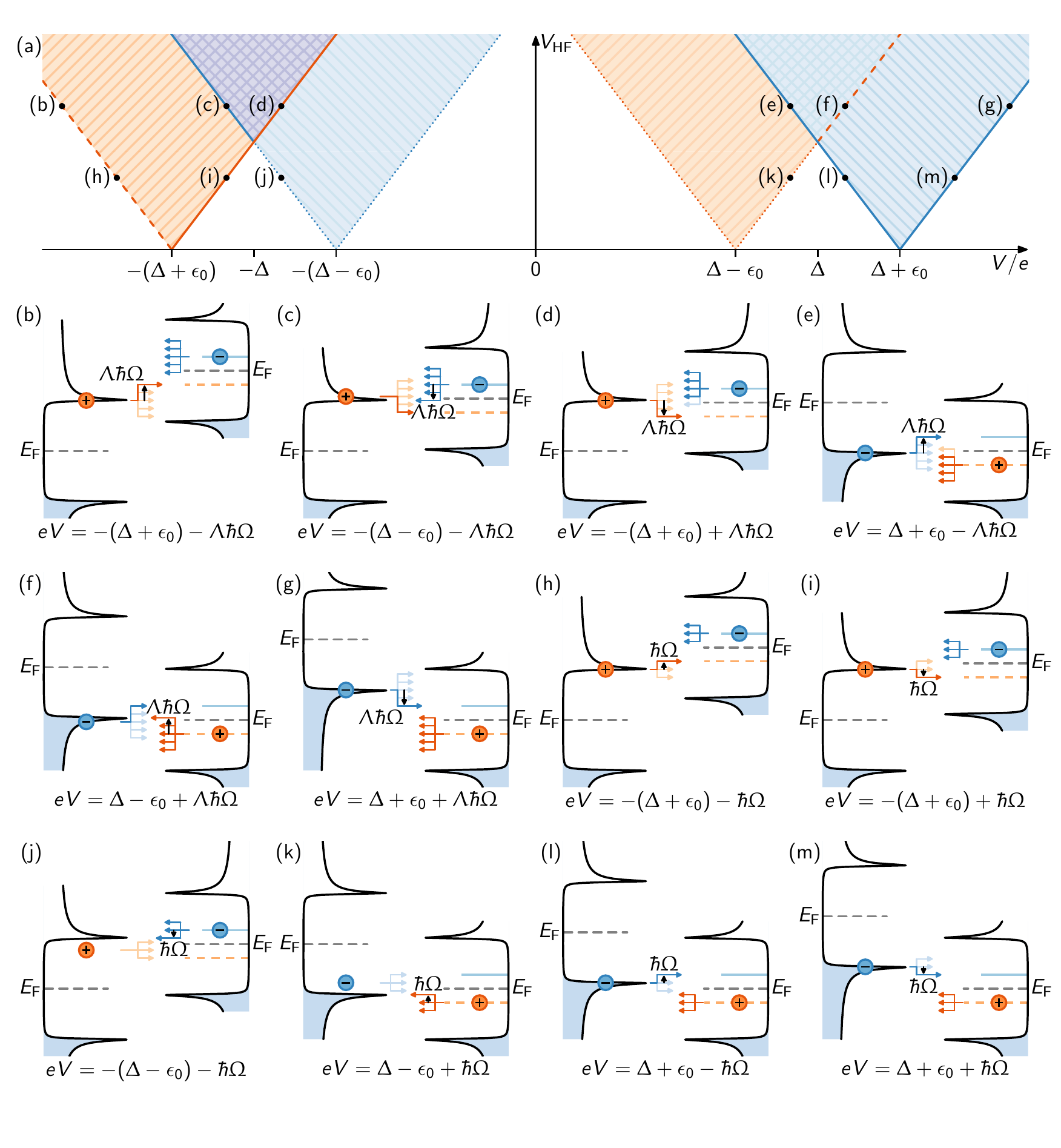} 
	\caption{\label{fig:OverviewOfTheProcesses} Portfolio of photon-assisted resonant Andreev tunneling processes illustrating various thresholds. (a) V-shaped regions with separate photon-assisted tunneling thresholds for electrons (blue; emerging from $V=\pm\Delta+\epsilon_0$) and holes (red; emerging from $V=\pm\Delta-\epsilon_0$). For resonant Andreev reflections to occur, both electron and hole tunneling must be above threshold, so that structure is only seen within outer V shapes. Due to a large asymmetry between electron and hole YSR wavefunctions at the tip position, electron tunneling is effectively rate limiting, and photon sidebands are most visible in the (blue) electron regions. At negative bias, this highlights the purple region, resulting in a pronounced Y shape, as indicated by full lines. At positive bias, the outer V (blue) is due to electron tunneling and sidebands are visible throughout. Dots marked (b)-(m) indicate bias voltages and HF amplitude for which tunneling processes are sketched in corresponding panels. Long arrows mark on-threshold processes, short full (faint) arrows mark above (below) threshold processes.}
\end{figure}

\Figure{PeakForPeak}b shows data for a high-conductance junction (small tip-substrate distance), so that tunneling into the YSR states is dominated by two-electron tunneling via resonant Andreev reflections (see \Fig{Thermal_Gamma}a for a sketch and discussion above). This tunneling process can be distinguished from single-electron tunneling on the basis of the dependence on the normal-state junction conductance as explained above (see further discussion in Sec.\ \ref{IntensityCrossing} below). Resonant Andreev reflection is described theoretically by the expression given in Eq.\ (\ref{eq:Ia}). 

At high junction conductance, we observe additional peaks within the tip gap. The peak at zero bias originates from Josephson tunneling. Multiple Andreev reflections (MAR) through the Pb--Pb junction can be found at $\pm 2\Delta/n$ for $n=2,3,...$. One also expects that resonant multiple Andreev reflections involving the YSR state appear at $\pm(\Delta+\epsilon_0)/n$ \cite{SFarinacci2018}. It is difficult to attribute the peaks labeled as MAR to a specific process due to the multitude of possible resonant Andreev processes associated to the three YSR states as well as (multiple) non-resonant MARs. The corresponding photon-assisted sidebands also remain unresolved.

\subsubsection{Tunneling processes in the presence of HF radiation}

Resonant Andreev reflections in the presence of HF irradiation involve photon emission and absorption for both electron and hole. Extending the processes shown in Fig.\ 4 of the main text, we provide a more complete portfolio of processes in \Fig{OverviewOfTheProcesses}. These processes demonstrate that there are independent thresholds for electron and hole tunneling, and that these thresholds can be relevant in overlapping regions in the $V-V_{\mathrm{HF}}$ plane, see the overlap regions of the two V shapes centered at $\pm(\Delta+\epsilon_0)$ and $\pm(\Delta-\epsilon_0)$. 

\begin{figure}[th]
	\includegraphics[]{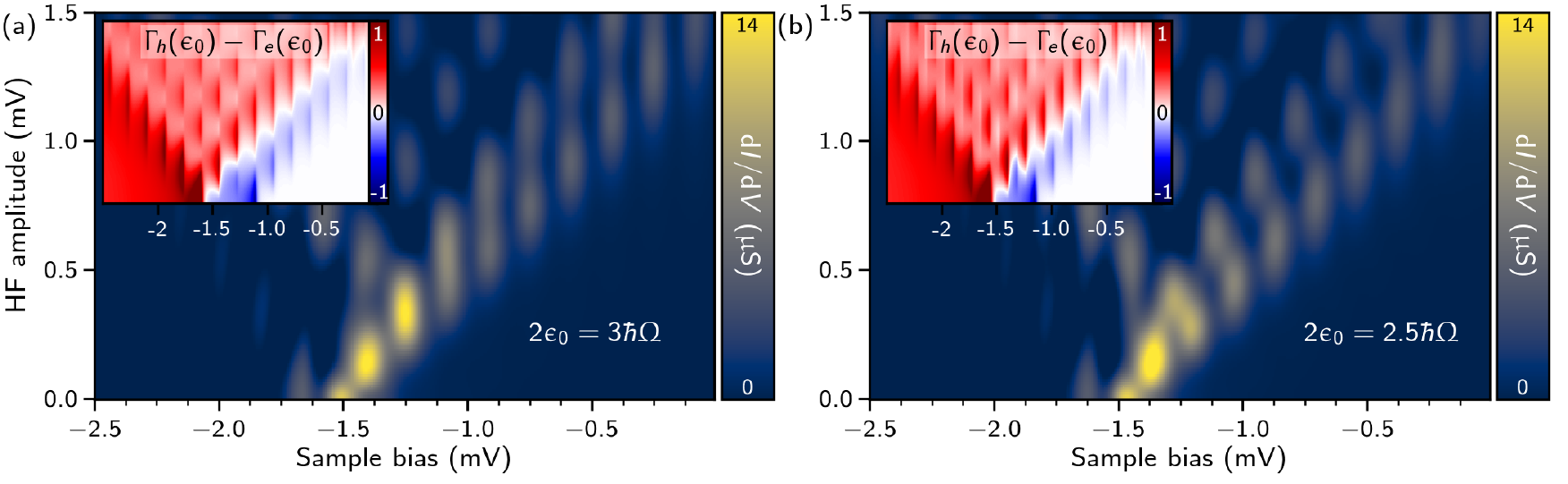}
	\caption{Simulated \didv spectra at negative bias of a YSR state under HF irradiation. Parameters are chosen to reproduce the Y shape observed in experiment. The peaks of the low voltage arm of the Y shape correspond to hole thresholds while the ones immediately above correspond to electron thresholds. In panel (a), we choose $2\epsilon_0=3\hbar\Omega$, where the peaks of both branches occur at the same sample bias. This situation is close to the actual experimental parameters. In panel (b), we choose $2\epsilon_0=2.5\hbar\Omega$, where the electron and hole thresholds are shifted by $\hbar\Omega/2$ from one another. The insets show the difference between the hole and electron tunneling rates in arbitrary units. Hole threshold peaks are only visible in the region were electron and hole tunneling rates are comparable and both nonzero, i.e., the white region between the blue and red areas in the insets. \label{fig:double}}
\end{figure}

\Figure{double} shows theoretical simulations on the double structure at the low-voltage arm of the Y pattern observed at negative biases for high junction conductance. For most parameters, the large asymmetry between hole and electron wave functions of the YSR state implies that only the electron thresholds for photon-assisted tunneling are visible in experiment. The underlying reason is that electron tunneling is weaker due to the smaller electron wavefunction, and thus effectively the rate-limiting process. The only exception occurs at the low-voltage side of the V-shaped region for hole tunneling at negative bias voltages. Here, the rate for electron tunneling, with threshods within the inner V shape, is already large, while hole tunneling is still weak since it is just barely allowed energetically. Thus, one expects to observe additional hole tunneling thresholds in the voltage range, where electron and hole tunneling are of the same order. The insets in \Figure{double} show the difference of these tunneling rates. Indeed, the double structure of the low-voltage arm of the Y pattern occurs just in the region where the difference between the tunneling rates becomes small. Panels (a) and (b) differ in the ratio between $2\epsilon_0$ and $\hbar\Omega$. For the parameters of panel (a), electron and hole thresholds appear at the same bias voltage. This is close to the situation in our experiment. In contrast, the two kinds of thresholds are shifted with respect to one another for the parameters in panel (b). Note that the lower structure (associated with hole tunneling) has a smaller apparent slope than the structure just above emerging from electron thresholds. The same apparent difference in slope is also clearly seen in the experimental data included in Fig.\ 3b of the main text.

\subsection{Simulation parameters}
\label{IntensityCrossing}

We emphasize that we can simulate the observed spectra for both single-electron and resonant Andreev tunneling via the YSR state based on a few parameters which can be extracted from experiment. This is distinctly different from the simulations in \Fig{ExperimentVsSimulation} which use the entire data set in the absence of HF irradiation as input. We now detail how we extract the necessary parameters. 

With the exception of the overall tunneling strength at large junction conductance, we extract all parameters from the \didv curves for low junction conductance in the absence of HF radiation. We assume that tip and substrate have not only the same gap, but also the same normal state density of states $\nu_0$. We can then extract the dimensionless product $\nu_0 t^\text{low}$ involving the tunneling amplitude $t^\text{low}$ from the expression 
\begin{equation}
	G_\text{N}= 4 \pi^2G_0 (\nu_0 t)^2
\label{eq:GN}
\end{equation}
for the normal-state differential conductance of a tunnel junction. This yields $\nu_0 t^{\text{low}} = \SI{8.1e-4}{}$. We can now choose parameters such that the peak heights and widths of the \didv curves in the absence of HF radiation are well reproduced by Eq.\ (\ref{eq:Is}). This can be used to extract $|u|^2/\nu_0$ from the single-electron tunneling at positive bias voltages and $|v|^2/\nu_0$ at negative bias voltages. Moreover, the thermal peaks depend sensitively on the relaxation rates $\Gamma_1$ and $\Gamma_2$. We also include a Dynes parameter to account for depairing interactions on the superconducting density of states \eqref{eq:SC_DOS_Theory},
\begin{equation}
	\label{eq:SC_DOS_Experiment}
   \nu(E)= \nu_0 \Re\left(\frac{E-\iu\Gamma_\text{S}}{\sqrt{(E-\iu\Gamma_\text{S})^2-\Delta^2}}\right)\, ,
\end{equation}
adding an imaginary part $\Gamma_\text{S}$ to the energy. In addition, we include a Gaussian averaging of the \didv traces with width $\Gamma_\text{broadening}=\SI{60}{\micro\electronvolt}$ which accounts for instrumental broadening. We find that the following set of parameters reproduces the experimental \didv curves: 
\begin{eqnarray}
   |u|^2/\nu_0 &=& \SI{0.21}{meV} \\
   |v|^2/\nu_0 &=& \SI{0.83}{meV} \\
   \Gamma_1 &=& \SI{0.70}{\micro\electronvolt} \\
   \Gamma_2 &=& \SI{0.11}{\micro\electronvolt} \\
   \Gamma_\text{S} &=&\SI{20}{\micro\electronvolt}  \\
   \Gamma_\text{broadening}&=&\SI{60}{\micro\electronvolt}  
\end{eqnarray}
In particular, we extract the ratio $|u|^2/|v|^2=0.253$.  

With this set of parameters, we can then simulate the \didv curves in the presence of the HF radiation, as shown in Fig.\ 3c of the main paper. 

Simulations of photon-assisted tunneling at high junction conductance require only one additional parameter, namely $\nu_0 t^{\text{high}}$. We choose this parameter so that we find good overall agreement between Eq.\ (\ref{eq:Ia}) and the measured data for high junction conductance in the absence of the HF radiation. With this procedure, we find $\nu_0 t^{\text{high}} = 39\times  \SI{8.1e-4}{}$. This is within 15\% of the value which one would extract from the junction conductance on the basis of Eq.\ (\ref{eq:GN}). This  completes the parameter set on which the simulation results shown in Fig.\ 3d of the main text are based.   

We note that this parameter set also allows one to corroborate that electron and hole tunneling is slow (fast) compared to the relaxation rates $\Gamma_1$ and $\Gamma_2$ for low (high) junction conductance, implying that there is a crossover between single-electron and resonant Andreev tunneling. Indeed, we find for the electron and hole tunneling rates, evaluated at the peak of the superconducting density of states (including depairing parameter), the values
\begin{eqnarray} 
   \Gamma_e &=& \SI{4.0}{\nano\electronvolt} \\
   \Gamma_h &=& \SI{15.8}{\nano\electronvolt} 
\end{eqnarray}
at low junction conductance, and 
\begin{eqnarray} 
   \Gamma_e &=& \SI{6.2}{\micro\electronvolt} \\
   \Gamma_h &=& \SI{24.5}{\micro\electronvolt} 
\end{eqnarray}
at large junction conductance. While the first set of tunneling rates are clearly smaller than $\Gamma_1$ and $\Gamma_2$, the second are significantly larger. 

\begin{figure}[tb]
	\includegraphics[]{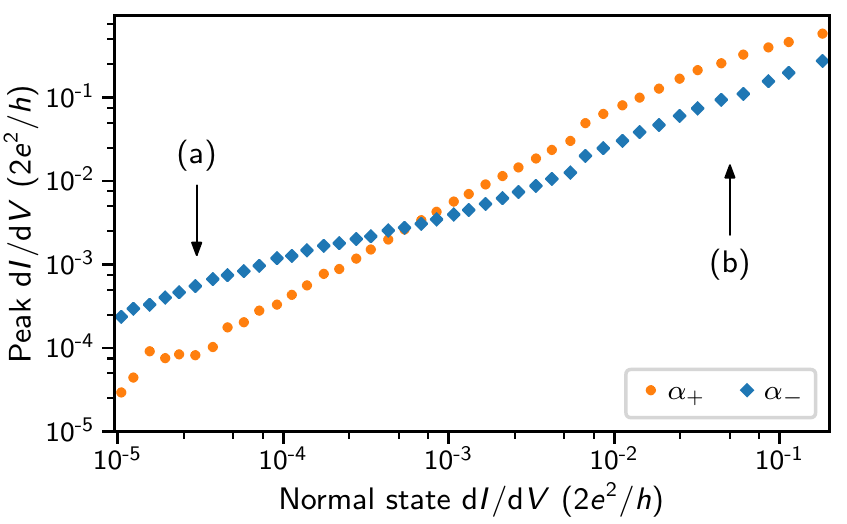} 
	\caption{\label{fig:S4} Peak heights $\alpha_\pm$ of the two resonances associated with the energetically lowest YSR state as a function of normal-state conductance at $T=\SI{1.35}{K}$. The labels (a) and (b) mark the conductance values at which the spectra in the corresponding panels of \Fig{PeakForPeak} were taken. 
	}
\end{figure}

This can also be double checked as follows. As a result of the asymmetry in the electron and hole wavefunctions, the amplitudes $\alpha_\pm$ of peak heights for tunneling into the YSR state with $\epsilon_0=\SI{0.25}{meV}$ reverses as the tip approaches the substrate and the normal state conductance of the junction increases \cite{SRuby2015b}. This is shown in \Fig{S4}. In particular, this indicates that panels (a) and (b) of \Fig{PeakForPeak} exhibit data in the regimes of dominant single-electron and dominant Andreev tunneling, respectively (see also the arrows in the figure).

\bibliography{PATunneling}

\end{document}